\documentclass[prd,amsmath,floats,amssymb, floatfix,superscriptaddress,nofootinbib,twocolumn]{revtex4} 

\usepackage[plainpages=false, colorlinks=true, anchorcolor=blue, linkcolor=blue, citecolor=blue, bookmarks=false]{hyperref}
\include{notations}
\usepackage[T1]{fontenc}
\usepackage{graphicx}% Include figure files
\usepackage{dcolumn}% Align table columns on decimal point
\usepackage{bm}% bold math
\usepackage{longtable}
\newcommand{\rthis}[1]{\textcolor{black}{#1}}
\usepackage{adjustbox}
\usepackage{float}
\usepackage[caption=false]{subfig}
\usepackage[paperwidth=210mm,paperheight=297mm,centering,hmargin=2cm,vmargin=2.5cm]{geometry}

\begin{document}

\preprint{APS/123-QED}

\title{Search for  GeV Gamma-Ray Emission from SPT-SZ selected Galaxy Clusters with 15 years of Fermi-LAT data}% Force line breaks with \\
%\thanks{A footnote to the article title}%

\author{Siddhant Manna}
 \altaffiliation{Email:ph22resch11006@iith.ac.in}%Lines break automatically or can be forced with \\
\author{Shantanu Desai}
 \altaffiliation{Email:shntn05@gmail.com}
\affiliation{
 Department of Physics, IIT Hyderabad Kandi, Telangana 502284,  India}

%\collaboration{MUSO Collaboration}%\noaffiliation

%\author{Charlie Author}
% \homepage{http://www.Second.institution.edu/~Charlie.Author}
%\affiliation{
 %Second institution and/or address\\
 %This line break forced% with \\
%}%
%\affiliation{
 %}%
%\author{Delta Author}
%\affiliation{%
 %Authors' institution and/or address\\
 %This line break forced with \textbackslash\textbackslash
%}%

%\collaboration{CLEO Collaboration}%\noaffiliation

%\date{\today}% It is always \today, today,
             %  but any date may be explicitly specified

\begin{abstract}
Galaxy clusters could produce gamma-rays from inverse Compton scattering of cosmic ray electrons  or hadronic interactions of cosmic ray protons with the intracluster medium. It is still an open question on whether gamma-ray emission ($>$ GeV energies)   has been detected from galaxy clusters. We carry out a systematic search for gamma-ray mission based on  300 galaxy clusters selected from  the 2500 deg.$^2$ SPT-SZ  survey after sorting them in descending order of $M_{500}/z^2$, using about 15 years of Fermi-LAT data in the energy range between 1-300 GeV. We were able to detect  gamma-ray emission with significance of about  $6.1\sigma$  from one  cluster, viz SPT-CL J2012-5649. The estimated photon energy flux from this cluster is approximately equal to $1.3 \times 10^{-6}$ MeV cm$^{-2}$ s$^{-1}$. The gamma-ray signal  is  observed between $1-10$ GeV  with the best-fit spectral index equal to $-3.61 \pm 0.33$.  However, since there are   six radio galaxies  spatially coincident with SPT-CL J2012-5649 within the Fermi-LAT PSF, we cannot rule out the possibility this signal could be caused by some of these radio galaxies.
Six other SPT-SZ clusters show evidence for gamma-ray emission with significance between $3-5\sigma$. 
None of the remaining  clusters show  statistically significant evidence for  gamma-ray emission.

\end{abstract}


\maketitle
%\tableofcontents
\section{\label{sec:level1}Introduction\protect}
Galaxy clusters are formed from the gravitational collapse of overdense regions in the early universe.  As the universe evolves, the   overdense regions created from density perturbations  accumulate more matter due to gravity, forming clumps and filaments that eventually merge to form clusters~\cite{Rees78}.  Therefore, galaxy clusters  constitute  the largest gravitationally bound and virialized structures in the Universe and act as a unique laboratory to probe cosmology~\cite{Kravtsov2012, Allen2011, Vikhlininrev} and fundamental Physics~\cite{Bohringer16,Desai18, Boraalpha, BoraDesaiCDDR, BoraDM}. Galaxy clusters have been observed over an extended wavelength range  from  radio waves~\cite{Feretti} to hard X-rays~\cite{Wik14}.  Over the past two decades, a large number of  dedicated surveys in optical, X-ray, and microwave have discovered  many new galaxy clusters, which have been used for a plethora of  Cosmology and Astrophysics studies, sometimes using a combination of observations at multiple wavelengths.  However, at higher energies  ($E > 1$~MeV), it is  still an open question, as  to  whether gamma-rays have been observed  from galaxy clusters.  This work is focused on searching for gamma-ray emission from galaxy clusters using  a mass limited catalog.

A number of mechanisms have been proposed for the production of  gamma-rays within  clusters, which we briefly recap.
Galaxy clusters constitute high concentrations of galaxies, dark matter (about 80\%), and hot diffuse gas (10-15\%). They are also giant reservoirs of high energy relativistic cosmic rays (CRs), i.e. relativistic electrons and protons swarming in the hot ionized Intra-Cluster Medium (ICM)~\cite{Brunetti2014,Wittor}. 
%Supernova driven galactic winds, direct injection in the cluster media through Active-Galactic Nuclei (AGN) outbursts, large-scale intergalactic shock waves generated during accretion and merger propagating in the ICM and turbulent re-acceleration have been proposed as some of the underlying mechanisms behind the acceleration of non-thermal CRs to high speeds~\cite{Huber13,Brunetti01,Shen21}.
Evidence for the acceleration of cosmic ray electrons comes from the observations of radio relics within clusters~\cite{Paul23,Wittor}. These relics result from  the shock waves generated during cluster mergers, accelerating particles  to extreme energies. These accelerated particles could produce gamma rays through Inverse Compton Scattering of relativistic electrons with the CMB,  non-thermal bremsstrahlung, or through the  decay of neutral pions produced from the collisions of cosmic ray  protons with the intracluster medium (ICM)~\cite{Ensslin97,Hinton07,2010PinkzeMNRAS.409..449P,Vazza16,Petrosian01,Brunetti17,2011BRUMNRAS.410..127B,Saqib2}.  \rthis{Most recently, the integrated gamma-ray flux from galaxy clusters  has been calculated by combining cosmological  MHD simulations of clusters of galaxies with the  cosmic ray propagation in the redshift range $z\leq 5$, where it was shown that this integrated flux could contribute up to 100\% of the diffuse gamma-ray flux   $z\leq 5$~\cite{Saqib}.}

%Other possible sources for the production of gamma rays include high-energy CR electrons in the AGN jets which can collide with low-energy photons and transfer some of their energy to the photons, boosting their energy to gamma-ray levels, CR protons in the AGN jet can also collide with gas in the intracluster medium (ICM), producing secondary particles, including pions. These pions can then decay into gamma rays~\cite{Ensslin97,Aharonian02,Hinton07}. When high-energy electrons in the SNR shock wave interact with nuclei in the interstellar medium, they can emit bremsstrahlung radiation. This radiation can be very energetic, extending into the gamma-ray regime~\cite{2010PinkzeMNRAS.409..449P, Petrosian01,Brunetti17,2011BRUMNRAS.410..127B}. 
Since most of the mass in galaxy clusters  is made up of non-baryonic cold dark matter, one could also detect  gamma rays  in clusters through the  annihilation of   dark matter WIMPs~\cite{Ackermann10,DiMauro23,Bringmann12,Lisanti,Murase23}.  Besides the aforementioned mechanisms for gamma-ray emission from the ICM, one could also 
obtain gamma-ray emission from star formation activity in cluster member galaxies~\cite{Storm}.

Before the launch of the Fermi Gamma-ray Space Telescope, the most definitive result on gamma-ray emission from galaxy clusters was reported in ~\cite{Reimer} using nine years of EGRET data from 1991-2000. This work reported upper limits for 58 X-ray-selected galaxy clusters  for energies between 100 MeV - 30 GeV.
In June 2008, NASA launched the Fermi Gamma-ray Space Telescope. The  Large Area Telescope (LAT) is one of the two instruments onboard this detector. Fermi-LAT is sensitive to high energy gamma rays from various astrophysical sources. It is a pair-conversion telescope that is sensitive to photons  between the energy range of 20 MeV to more than 300 GeV~\cite{Atwood09}. A plethora  of studies have used the Fermi-LAT data to look for both diffuse  broadband~\cite{2010Ackermann...717L..71A,Arlen12, 2012Han.427.1651H, Prokhorov14,ProkhorovMNRAS,2014Ackermann...787...18A,Dutson13,Griffin14,Zandanel14,FermiVIRGO,Ackermann16,Branchini17,Xi18} and line emissions from galaxy clusters~\cite{Liang16,Anderson16}.   We report  a few  salient highlights from some of the above works.  Among all the above searches,  no extended  broad-band gamma-ray emission has been unambiguously detected from the cluster ICM, except for the Coma cluster. For all other clusters, any  putative gamma-ray signal seen in searches from galaxy clusters has been attributed to   AGNs (such as  blazars) located inside the cluster~\cite{Abdo09,Abdo2,Prokhorov14}. A large number of works have looked for gamma-ray emission from the  Coma cluster (Abell 1656) with Fermi-LAT. The discovery of a massive radio halo and radio remnants suggests efficient particle acceleration in the Coma Cluster~\cite{Giovannini93}. Although initial searches by the Fermi-LAT Collaboration as well as by other authors  found no statistically significant gamma-ray emission from the Coma cluster~\cite{Ackermann10,Ackermann16,Zandanel14,ProkhorovMNRAS}, other works have found statistically significant emission from the Coma cluster with accumulated livetime. In 2017,  a ring-like structure on the fringes of the Coma galaxy cluster was discovered using eight years of Fermi-LAT data with 3.4$\sigma$ significance~\cite{Reisscoma}. This detection was confirmed in \cite{Xi18}  using nine years of Fermi-LAT data with the observed significance $> 5\sigma$, and reaffirmed in ~\cite{Baghmanyan22},  who found extended diffuse gamma-ray emission with 5.4$\sigma$ significance based on  12.3 years of  data. The theoretical implications of this detection from Coma cluster  are discussed in ~\cite{Adam21}. No significant emission was seen from the VIRGO cluster, although emission was detected from two elliptical galaxies, M87 and M49 located near the VIRGO center~\cite{FermiVIRGO}.   In addition to searches from the ICM,  a search from  114  Brightest cluster galaxies (BCGs) selected  from multiple X-ray catalogs, containing radio sources with flux above 50/75 mJy was done using 45 months of Fermi-LAT data~\cite{Dutson13}. This search detected signals from  four possible  sources, although none  of them could be  unambiguously associated with the BCGs~\cite{Dutson13}.

In addition to the above pointed searches from individual clusters, many works  have also  done a stacking analysis from multiple clusters. The first such study looked for stacked emission from 53 clusters in the HIFLUGCS sample~\cite{Chen07} and did not detect any  significant emission~\cite{Huber13}. A similar search was done using  the Fermi-LAT data above 10 GeV by stacking 55 clusters from the HIFLUGCS sample~\cite{Prokhorov14}. A $4.3\sigma$ excess was obtained from this analysis, which was attributed to contribution from AGNs~\cite{Prokhorov14}. A similar stacking analysis of the Fermi-LAT data using  78 clusters ($z<0.12$) from the 2MASS survey reported null results~\cite{Griffin14}.  Another stacked search using 112 clusters in the MCXC catalogue found evidence at $5.8\sigma$ significance for  a central point source dominated by AGN emission along with a gamma-ray ring at the position of the virial shock~\cite{Reiss18}.  

In addition to  the aforementioned pointed  or stacked searches from X-ray selected cluster samples, a novel search for gamma-ray emission was done by cross-correlating the cluster positions from   SDSS and  Planck selected  clusters with the Fermi-LAT data, and calculating the two-point correlation function~\cite{Branchini17}. A positive correlation was seen from this search, attributed to cumulative emissions from AGNs. However, a definitive conclusion as to whether the cross-correlation is because of AGNs inside the cluster or diffuse emission within the ICM could not be drawn~\cite{Branchini17}.

Motivated by some of the above works, which found tantalizing hints for gamma-ray emission from clusters, we would like to systematically search the Fermi-LAT data  using galaxy clusters detected using the Sunyaev-Zel'dovich (SZ) effect~\cite{SZ}. The SZ effect arises from  the interaction of CMB photons with high energy electrons in  galaxy clusters through Inverse Compton Scattering. The SZ effect is sensitive to the cluster mass threshold and is independent of redshift. Therefore SZ surveys provide a mass-limited catalog independent of redshift~\cite{Birkinshaw,Holder}.

This manuscript is structured as follows. In Sect.~\ref{sec:level2}, we describe the SPT-SZ cluster sample used for our analysis.  In Sect.~\ref{sec:level3}, we explain the Fermi-LAT data analysis procedure used to search for gamma-ray emission.  Our  results are discussed  in Sect.~\ref{sec:level4}. Finally, our conclusions can be found in Sect.~\ref{sec:conclusions}. For our analysis, we assume a flat $\Lambda$CDM cosmology with $\Omega_m = 0.3$ and  $h=0.7$.

\section{\label{sec:level2}Cluster Selection\protect}
The dataset  used for our analysis consists of galaxy clusters detected by the  South Pole Telescope (SPT). The SPT  is a 10-meter telescope located at the South Pole that  has imaged the sky at three different frequencies:  95 GHz, 150 GHz, and 220 GHz~\cite{Carlstrom}. SPT completed  a 2500 square-degree survey between 2007 and 2011 to detect galaxy clusters using the SZ effect. This 2500 sq. degree  SPT-SZ survey detected 677 confirmed galaxy clusters with  SNR greater than 4.5, corresponding to  a  mass threshold of $3 \times 10^{14} M_{\odot}$ up to redshift of  1.8~\cite{Bleem15,Bocquet19}\footnote{{\url{ https://pole.uchicago.edu/public/data/sptsz-clusters/2500d_cluster_sample_Bocquet19.fits }}}. SPT has an angular resolution of  approximately 1 arcminute~\cite{Carlstrom}. The SPT cluster redshifts  have been obtained using a dedicated optical and infrared follow-up campaign, consisting of  pointed imaging  and spectroscopic observations~\cite{Song,Ruel}, as well  as using the data from optical surveys such as BCS~\cite{Desai12} and DES~\cite{Saro}.  The original SPT telescope has subsequently  been upgraded with new instrumentation and has conducted additional cluster surveys using SPTPol~\cite{Bleem20}, and in the future, will be superseded by SPT-3G~\cite{Benson14}. Similar to ~\cite{Huber13}, we carried out a search for gamma-ray emission from 300  clusters from the above sample in decreasing order of $M_{500}/z^{2}$, where  $M_{500}$ is the total mass contained within a sphere with an average density equal to 500 times the critical density of the universe at the cluster's redshift and  $z$ is the cluster's redshift~\cite{Bocquet19}.  \rthis{Note that this choice of ordering is arbitrary and we could have instead ordered it according to $M_{500}/D_L^{2}$.}
%The list of clusters analyzed with their masses, redshifts, and angular si can be found in Table~\ref{tab:TableI}.

\section{Fermi-LAT data analysis}
\label{sec:level3}
We used the data from the Fermi-LAT \texttt{Pass 8 ULTRACLEANVETO (`FRONT + BACK')} class events ~\cite{Atwood2013Pass8T} spanning almost 15 years of data (\texttt{MET 239587201-710640005}) from August 5, 2008 to July 10, 2023. This event class was chosen because it had the lowest amount of cosmic ray contamination, which makes it perfect for studying diffuse emissions. The data were chosen within a $5^{\circ}$ radius around the cluster center   (based on the SPT-derived position) between 1000 MeV and 300 GeV. Due to the large Point Spread Function (PSF) at lower energies, we avoided analyzing data  below 1 GeV \cite{Atwood2013Pass8T}. At the lowest energy of 1 GeV considered, the PSF is around $1.72^{\circ}$ and for the highest energy  considered of  300 GeV, the PSF is found to be $0.17^{\circ}$~\cite{Fermi12}.

We used the \texttt{Fermipy (version 1.2~\cite{Wood17})} and \texttt{Fermitools v2.2.0} software packages to analyze the data using the binned maximum-likelihood analysis technique with the \texttt{P8R3\_ULTRACLEANVETO\_V3} instrument response functions (IRFs). We used the {\tt Fermibottle} Docker container and analysis environment provided by the Fermi Science Support Center.\footnote{available at  \url{https://github.com/fermi-lat/FermiBottle}} The recommended \texttt{(DATA\_QUAL $>$ 0)} and \texttt{(LAT\_CONFIG == 1)} data quality filters were used for the data reduction. For better quality data and more refinement, we also applied \texttt{abs}(\texttt{rock angle}) $< 52^{\circ}$ and \texttt{$|b| < 20^{\circ}$} as additional cuts. To further reduce contamination from the Earth's atmosphere, we used a zenith angle cut of $90^{\circ}$ to the events. We used 0.2-pixel resolution for the spatial binning and 10 logarithmic energy bins per decade for the spectral binning in the energy range 1 GeV - 300 GeV.  
%To account for diffuse emission, we used the Galactic diffuse emission model \texttt{(gll\_iem\_v07.fits)} with an isotropic component \texttt{(iso\_P8R3\_ULTRACLEANVETO\_V3\_v1.txt)} appropriate to the \texttt{ULTRACLEANVETO} event class.
\subsection{Background Model}
In our background model, we included all the sources from  the fourth Fermi-LAT catalog of gamma-ray sources (4FGL-DR4), consisting of  both point-like and extended sources~\cite{Fermi23}. To account for the diffuse emission, we used the Galactic diffuse emission model \texttt{(gll\_iem\_v07.fits)} with an isotropic component \texttt{(iso\_P8R3\_ULTRACLEANVETO\_V3\_v1.txt)} appropriate to the \texttt{ULTRACLEANVETO} event class.
We allowed  the normalization of the templates used to describe the Galactic foreground and isotropic diffuse emission to vary. The Fermi-LAT background models are divided into the Galactic diffuse model and the isotropic spectral template. The Galactic diffuse model consists of a spatial and spectral template that describes the emission from the Milky Way. The isotropic spectral template provides the spectral form from a fit to the all-sky emission not represented in the Galactic diffuse model. 

\subsection{Binned Likelihood Analysis}
We used the conventional binned-likelihood analysis method outlined by the Fermi-LAT team to perform the likelihood analysis and model fitting.  The spectral parameters of sources inside the region of interest (ROI)  equal to $3^{\circ}$, were kept as free parameters during the fitting, whereas  those for 4FGL-DR4 sources outside the ROI  but within $5^{\circ}$ were kept as constant during the fit. For our analysis, we utilize a point-source template for the gamma-ray detection with $\Gamma=2$ similar to~\cite{Baghmanyan22}.  
We used the  software \texttt{gtselect} to select the events (gamma-ray photons) from the Fermi-LAT data based on multiple criteria such as the energy range, time span, ROI, and data quality. It enabled us to generate a filtered event file with the properties  needed for our investigation. Utilizing the \texttt{gtmktime} function, we then generated a time filter for the  Fermi-LAT data. This tool excluded time intervals with strong background activity, such as passes through the Earth's radiation belts or periods of sensor maintenance.  Then, we generated a ROI counts map, summed over the photon energies, to identify candidate sources and validate that the field looks reasonable as a basic sanity check. For this purpose,  we used the \texttt{gtbin} tool with the ``CMAP'' option. The data input for the binned likelihood analysis is a three-dimensional count map with an energy axis known as a counts cube. The counts cube is a square binned region displayed in the count's map that must fit within the circular acceptance cone defined during the data extraction.  To calculate the exposure, we use the livetime cube, which  is a three-dimensional array representing the time the LAT observed each position in the sky at each inclination angle and is required for precise flux and spectral analyses. It is created using the \texttt{gtltcube} software package. It adjusts for the exposure changes caused by the spacecraft's orbit and instrument livetime, and finally  computes this exposure time as a function of energy for a region of interest. 

Then, the source maps generated by \texttt{gtsrcmaps} are utilized for the likelihood  analysis. These maps depict the projected counts from all sources within a specific energy range and spatial region. 
Finally, we generate the model maps using \texttt{gtmodel}.  When the model closely aligns with the actual gamma-ray emissions in the observed region, the resulting model map should closely mirror the counts map.

\subsection{Model fitting using Maximum Likelihood Estimate}
We used Maximum Likelihood Estimate (MLE) to find the best-fit model parameters that describe the source’s spectrum and position~\cite{Mattox96}. We use \texttt{gtlike} to carry out  the binned likelihood analysis of  the LAT data. It works by taking a model of the gamma-ray sky and calculating the probability of observing the data given that model. The model includes information about the locations, spectra, and other source properties.  For source detection, we calculate the Test Statistic (TS) using \texttt{gttsmap} to characterize the significance of  gamma-ray sources, which is given as follows:
\begin{equation}
    TS = -2 \ln \left(\frac{L_{\text{max},0}}{L_{\text{max},1}}\right),
\end{equation}
$L_{\text{max},0}$ corresponds to the null hypothesis (MLE for
the model without the signal model), and $L_{\text{max},1}$ is the alternative hypothesis (signal model at the specified location).  Wilks’ Theorem states that for large counts,  TS  for the null hypothesis is asymptotically distributed as a $\chi^2$ distribution~\cite{Wilks}.  The detection significance (or $Z$-score) is equal to  $\sqrt{TS}$. This  TS statistic is also widely used in neutrino astrophysics  to quantify the detection significance~\cite{Pasumarti}.

\section{\label{sec:level4}Results\protect}
\begin{figure*}[!t] % Use figure* instead of figure
  \centering
  \includegraphics[width=\textwidth]{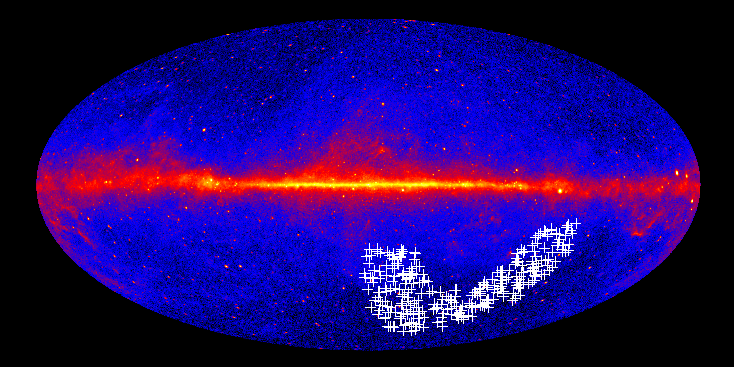} % Replace with your figure file
  \caption{The Fermi-LAT count map in galactic coordinates  based on the 4FGL-DR4 catalog, using  12 years of survey data~\cite{Fermi20,Fermi23}. The white colored  plus sign depicts the locations of the 300 SPT-SZ galaxy clusters we used  in our analysis. The bright, diffuse glow running along the middle of the map, shows  the central plane of our Milky Way galaxy.}
  \label{fig:Figure1}
\end{figure*}

We  implemented the aforementioned  MLE procedure on  300 SPT-SZ galaxy clusters sorted in  decreasing order of their $M_{500}/z^{2}$ values. These 300 clusters have been juxtaposed  on the 12 year  Fermi point-source catalogue  skymap in galactic coordinates  in Fig.~\ref{fig:Figure1}.~\footnote{For generating Fig.~\ref{fig:Figure1}, we  have used the FITS file available at  \url{https://fermi.gsfc.nasa.gov/ssc/data/access/lat/12yr_catalog/intens_scaled_ait_144m_gt1000_psf3_gal_0p1.fits.gz}.}   Our results from the MLE anlysis  for these clusters are tabulated  in Table~\ref{tab:TableI}. Each row contains the cluster $M_{500}$ in units of $(\times 10^{14} M_{\odot})/h$, redshift, RA, and declination, all of which were obtained from ~\cite{Bleem15} and finally the corresponding TS value. For this sample of clusters, the virial radius of the clusters corresponds to a mean subtended angle $\theta_{200} = 0.039^{\circ}$ and corresponding standard deviation equal to 0.026$^{\circ}$. 
We found only one cluster (SPT-CL J2012-5649)  with detection significance $> 5\sigma$ (TS=37.2). In addition, there are six clusters with detection significance greater than $3\sigma$.  
These clusters include  SPT-CL J2021-5257 (TS=12), SPT-CL J0217-5245 (TS=11.9), SPT-CL J0232-5257 (TS=11.5), SPT-CL J0619-5802 (TS=10.3), SPT-CL J0124-4301 (TS=9), and SPT-CL J2140-5727 (TS=9) respectively.  All other clusters had TS values $< 9$  (or $<3\sigma$ significance). 

We note two noteworthy merging clusters in our sample: the Bullet Cluster (SPT-CL J0658-5556) and El Gordo Cluster (SPT-CL J0102-4915). These have TS values of 2.1 and 4.7, respectively, corresponding to no significant excess. The Bullet cluster is a merging cluster that has been used to test $\Lambda$CDM and modified gravity theories~\cite{Clowe06}. The gamma-ray luminosity for the Bullet cluster has also  been previously estimated  in literature from the observed infrared luminosity, but was shown to be undetectable by Fermi-LAT due to its large redshift~\cite{Storm}.  The Fermi-LAT collaboration also did not find evidence for gamma-ray emission from the Bullet cluster using  the first 1.5 years of data~\cite{2010Ackermann...717L..71A}.
The El Gordo cluster is one of the most massive merging  galaxy clusters ($M_{200} \sim 3 \times 10^{15} M_{\odot}$) located at a redshift of 0.87~\cite{Menanteau} which has also been extensively used to test $\Lambda$CDM~\cite{Kroupa}.
Therefore, there is no evidence for gamma-ray emission from these two merging clusters using 15 years of data.

We now discuss the results for the  clusters with observed detection significance $> 3\sigma$. We first focus on  SPT-CL J2012-5649 in detail, followed by the other clusters.

%Please fix rest of sentence above

%this paragraph shoudl be AFTER you discuss J2012-5649

\subsection{SPT-CL J2012-5649}
We found a distinctive gamma-ray emission signature for SPT-CL J2012-5649. This cluster is located at a redshift of 0.055,    SPT S/N ratio of 5.99, corresponding to $M_{500} \approx  5 \times 10^{14} M_{\odot}$ and angular diameter given by $\theta_{200}= 0.19^{\circ}$.

\begin{figure}
\centering
\begin{adjustbox}{right=1\columnwidth}
\includegraphics[width=1\columnwidth]{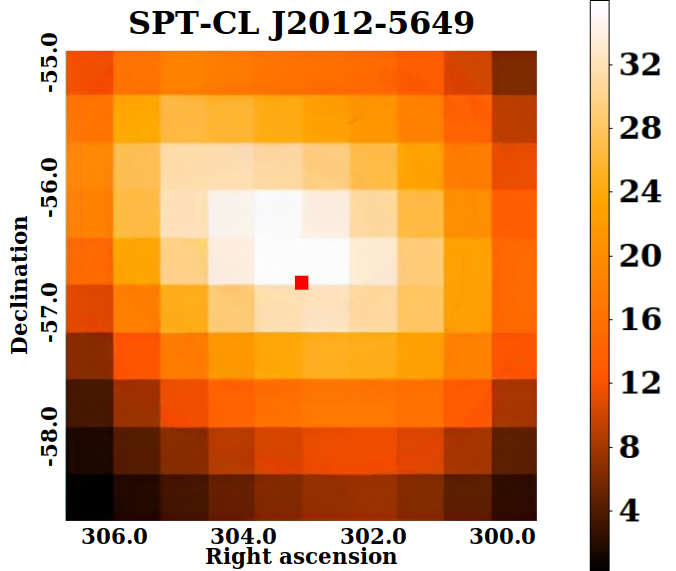}
\end{adjustbox}
\caption{Gaussian kernel smoothed ($\sigma = 1.5$) TS map of the SPT-CL J2012-5649 cluster (left) and TS map scale (right) generated using \texttt{gttsmap} in the energy band 
$1- 300$ GeV. We used 0.2-pixel resolution for the spatial binning. The red square shows the SPT cluster center.}
\label{fig:TS2012}
\end{figure}

\begin{figure}
\centering
\begin{adjustbox}{right=1\columnwidth}
\includegraphics[width=1\columnwidth]{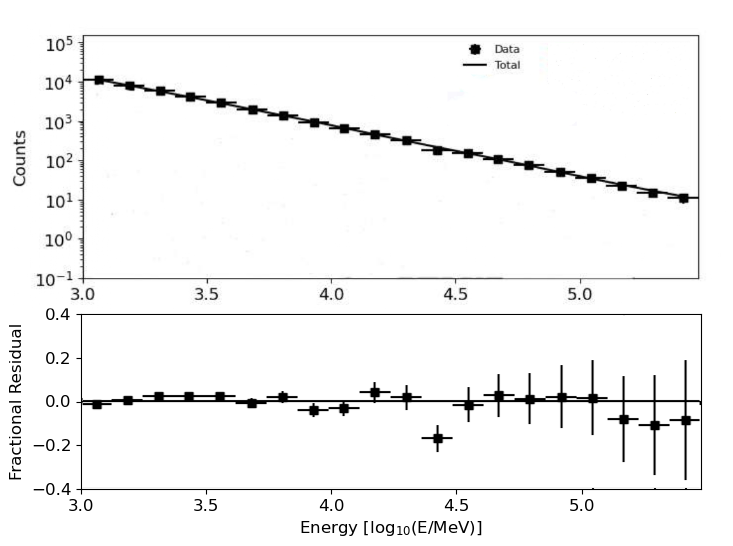}
\end{adjustbox}
\caption{Top: Observed photons in energy bins from 1-300 GeV and the  cumulative model of the total emission   all the 4FGL-DR4 source diffuse  emission templates and the observed signal from SPT-CL J2012-5649.  Bottom: The fractional residuals given by (counts-model)/model, determined within  $5^{\circ}$ for SPT-CL J2012-5649.}
\label{fig:figure3}
\end{figure}

\begin{figure}
\centering
\begin{adjustbox}{right=1.05\columnwidth}
\includegraphics[width=1.2\columnwidth]{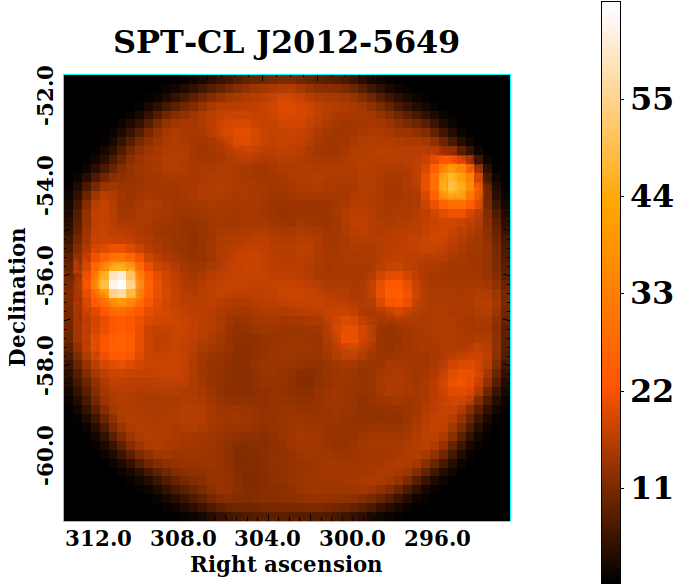}
\end{adjustbox}
\caption{Gaussian kernel smoothed ($\sigma = 1.5$) counts map of SPT-CL J2012-5649 cluster (left) generated using \texttt{gttsmap} in the energy range $1-300$ GeV. We used 0.2-pixel resolution for the  spatial binning.}
\label{fig:countmap2012}
\end{figure}

\begin{figure}
\centering
\begin{adjustbox}{right=1\columnwidth}
\includegraphics[width=1\columnwidth]{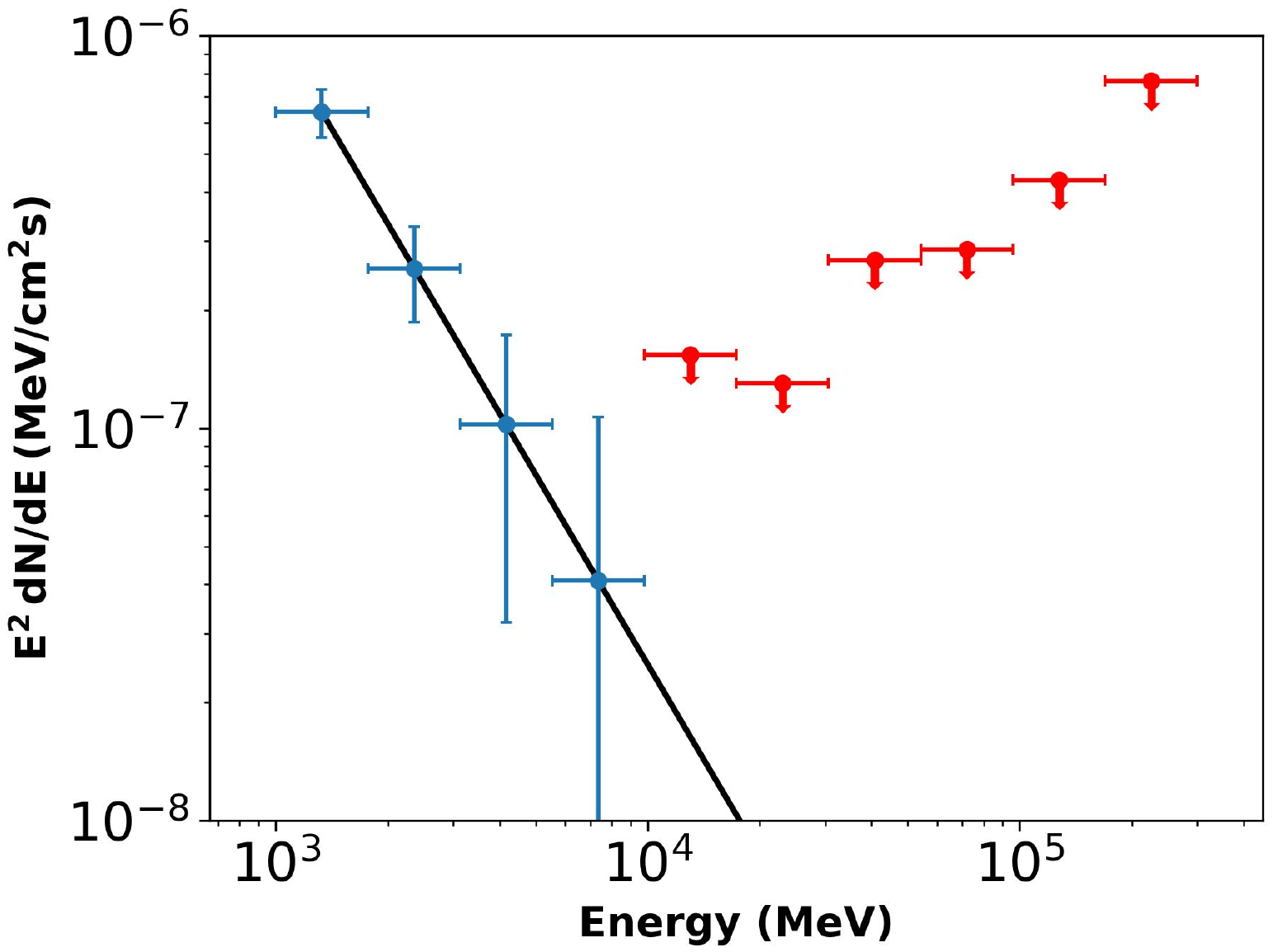}
\end{adjustbox}
\caption{The differential energy spectrum from SPT-CL J2012-5649 obtained using {\tt easyFermi}. The solid line shows the power-law fit with  the best-fit spectral index given by $\gamma = -3.61 \pm 0.329$. The blue data points show  the measured differential energy spectrum, while the red point represent  upper limits. All the signal is observed at energies $<=10$ GeV, and beyond that we obtain upper limits.}
\label{fig:SEDplot}
\end{figure}

The TS  map  for SPT-CL J2012-5649 using a power law point source template with the photon spectral index $\Gamma=-2$,  after smoothening with a Gaussian kernel ($\sigma=1.5$) can be found in  Fig.~\ref{fig:TS2012}.
This cluster has a  TS value of around 37.2, corresponding to a detection significance of  $6.1\sigma$ for this point source template.   The observed signal is confined to about $0.2 R_{200}$.  We found no extended sources in our search.
We also show the count maps and residuals in Fig.~\ref{fig:figure3}.
The top panel  shows the observed photons in the energy bins from 1-300 GeV along with the total signal, given by the sum of  4FGL-DR4 sources, galactic diffuse emission templates, and the observed emission from SPT-CL J2012-5649,  whereas the fractional residuals determined from the full ROI of  $5^{\circ}$ for SPT-CL J2012-5649 are shown in the bottom panel.  In Fig.~\ref{fig:countmap2012}, we depict the Gaussian kernel smoothened count map ($\sigma = 1.5$).  The total  photon flux for this cluster is equal to  $(0.39 \pm 0.05) \times 10^{-9}$ ph cm$^{-2}$ s$^{-1}$ and the total energy flux corresponds to  $(0.63 \pm 0.09) \times 10^{-6}$ MeV cm$^{-2}$ s$^{-1}$.  In Fig.~\ref{fig:SEDplot}, we showcase the observed  Spectral Energy Distribution (SED)  for SPT-CL J2012-5649 along with the best-fit spectrum. We found the best-fit  spectral index ($\gamma$)  given by  $\gamma = -3.61 \pm 0.33$, where $\frac{dN}{dE} \propto E^{\gamma}$. All the observed signal is between 1-10 GeV. For energy $>$ 10 GeV, we only obtain  upper limits. \rthis{We also note that the fitted spectral index is independent of the choice of the spectral index used for the point source template.}

For this cluster, we also did a search with other templates. For a  radial disk and radial Gaussian  templates, we found the TS value around 5.61~($2.36\sigma$). We also did a search in the energy range between 10 GeV to 300 GeV, we found the TS value to decrease significantly to 6.0~($2.4\sigma$). This is in accord with the observed SED which does not show any detections  beyond 10 GeV.

\hfill

\hfill

\hfill

SPT-CL J2012-5649 is spatially coincident with Abell 3667.  Abell 3667 is also one of the most active galaxy clusters, with several ongoing mergers and collisions between galaxies. It is one of the brightest X-ray sources in the southern sky~\cite{Edge92}.  One of the most striking features of Abell 3667 is the large shock wave propagating through it. This shock wave was created when two smaller galaxy clusters collided and merged to form Abell 3667~\cite{deGaasperin22,Carretti13}. The shock wave   accelerates particles, creating a population of relativistic electrons emitting radio waves.  Two giant radio relics, that are thought to be caused from the acceleration of electrons by the shock wave  have also been detected within this cluster with ATCA and MeerKAT~\cite{deGaasperin22,Carretti13,Riseley}. \rthis{From the MeerKAT data, the presence of a pair of radio relics in the galaxy cluster Abell 3667 was reported~\cite{deGaasperin22}. The North Western (NW) radio relic has an angular size of \boldmath{$0.6^{\circ}$} and an extension of 2.3 Mpc. The South Eastern (SE) relic is comparatively smaller, with an angular size of \boldmath{$0.42^{\circ}$.} and an extension of 1.6 Mpc. Furthermore, ATCA data was used  to reveal the expansive nature of the NW radio relic, determining its total extent to be 2.6 Mpc~\cite{Rottgering1997}. This suggests that  their angular sizes are comparable with the PSF of Fermi-LAT. } It would be intriguing to correlate the gamma-ray flux with the radio flux and to redo our analysis using a template obtained from observations of this radio relic, similar to studies done for the Coma cluster~\cite{Xi18}. One could also estimate the observed gamma-ray flux for this cluster using the {\tt MINOT} software~\cite{MINOT}. These  analyses shall be deferred to future work.

We, however, caution that any possible gamma-ray signals from galaxy clusters could be due to contamination by radio galaxies and blazars within the clusters. Fermi-LAT has detected gamma rays from radio galaxies at the centers of VIRGO and Perseus clusters~\cite{Abdo09,Abdo2}.
Furthermore, the MeerKAT observations of this cluster have revealed three radio galaxies within six arcminutes of the cluster center~\cite{deGaasperin22}.
Therefore, to check for such a contribution,  we  searched for coincident radio sources  using the  Sydney University Molonglo Sky Survey (SUMSS) catalog~\cite{Mauch03}.   Fermi-LAT has also detected GeV emission from radio galaxies and AGNs, including from SUMSS sources~\cite{FermiAGN}. 
%We could not find spatial coincidences with sources in the SUMSS catalogue within 1 arcminute of the SPT cluster center.
Within $0.2 ^\circ$, we found six SUMSS radio sources: SUMSS J201156-564547 with 17.5 mJy integrated 36-cm flux density, SUMSS J201142-564759 with 47.3 mJy integrated 36-cm flux density, SUMSS J201127-564358 with 535.7 mJy integrated 36-cm flux density, SUMSS J201125-564312 with 495.1 mJy integrated 36-cm flux density, SUMSS J201113-565408 with 13.4 mJy integrated 36-cm flux density, SUMSS J201317-565906 with 8.3 mJy integrated 36-cm flux density.  These SUMSS sources have been classified as radio galaxies in SIMBAD,   and are  also within 10 arcminute from the three MeerKAT detected radio galaxies. 
%We also cross-matched with the  latest Fermi-LAT catalogue 4FGL-DR4 ~\cite{Fermi23} within $0.2^{\circ}$ and found no coincidences. % We also found spatial correlation with five of our SUMSS sources: SUMSS J201156-564547, SUMSS J201142-564759, SUMSS J201127-564358, SUMSS J201125-564312, and SUMSS J201317-565906 with the three Radio galaxies detected by \cite{deGaasperin22} within an angular separation of $10'$ using SIMBAD. 

Therefore,  our results  imply a $6\sigma$ detection of gamma-rays from this cluster, which prima-facie
is only the second cluster (after Coma) with statistical significance $> 5\sigma$.
Nevertheless, since we found six SUMSS radio sources within the Fermi-LAT PSF, we cannot arbitrate  as to whether the observed gamma-ray emission within this cluster is due  to physical processes  within the ICM or due to contamination  from these radio sources.
To the best of our knowledge, there has not been any previous result related to gamma-ray emission  for  this cluster using Fermi-LAT. However, a  search  for TeV gamma rays was done from Abell 3667 using the CANGAROO-III atmospheric Cherenkov telescope, which reported null results~\cite{Cangaroo}.

\subsection{Clusters with significance between $3-5\sigma$}
\begin{figure}
\centering
\begin{adjustbox}{right=1.0\columnwidth}
\includegraphics[width=1.0\columnwidth]{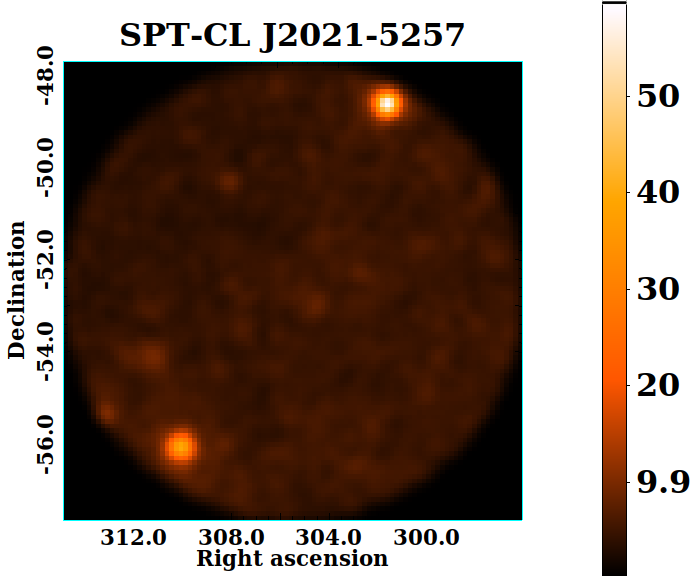}
\end{adjustbox}
\caption{Gaussian kernel smoothed counts map of the SPT-CL J2021-5257 cluster done in the same way as in Fig.~\ref{fig:countmap2012}.}
\label{fig:figure6}
\end{figure}

\begin{figure}
\centering
\begin{adjustbox}{right=1\columnwidth}
\includegraphics[width=1\columnwidth]{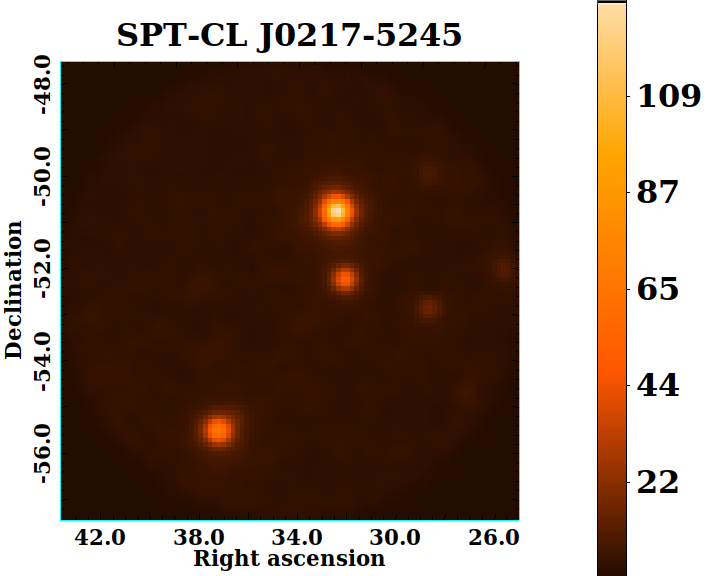}
\end{adjustbox}
\caption{Gaussian kernel smoothed counts map of the SPT-CL J0217-5245 cluster done in the same way as in Fig.~\ref{fig:countmap2012}.}
\label{fig:figure7}
\end{figure}

\begin{figure}
\centering
\begin{adjustbox}{right=1\columnwidth}
\includegraphics[width=1\columnwidth]{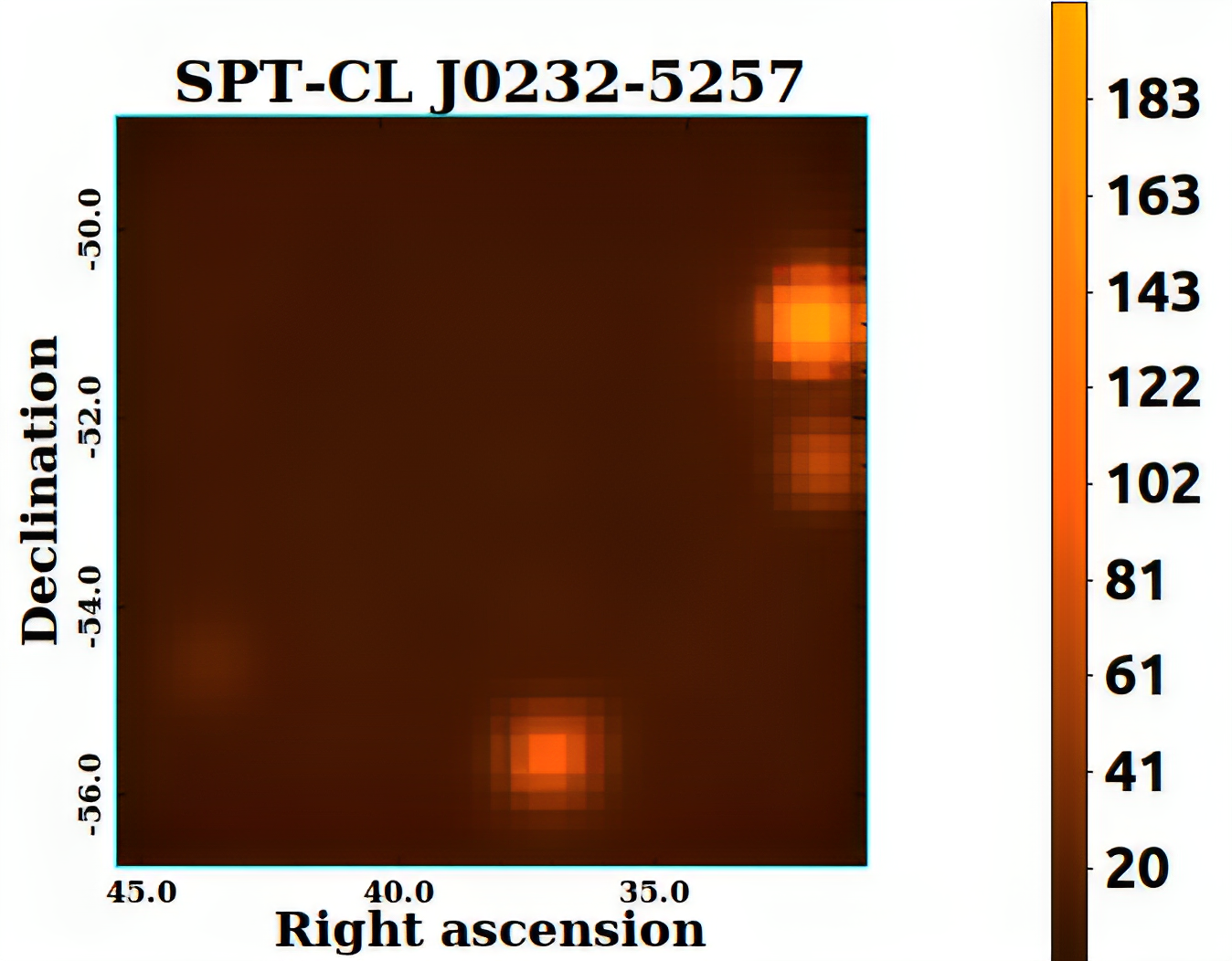}
\end{adjustbox}
\caption{Gaussian kernel smoothed counts map of the SPT-CL J0232-5257 cluster done in the same way as in Fig.~\ref{fig:countmap2012}.}
\label{fig:figure8}
\end{figure}

\begin{figure}
\centering
\begin{adjustbox}{right=1\columnwidth}
\includegraphics[width=1\columnwidth]{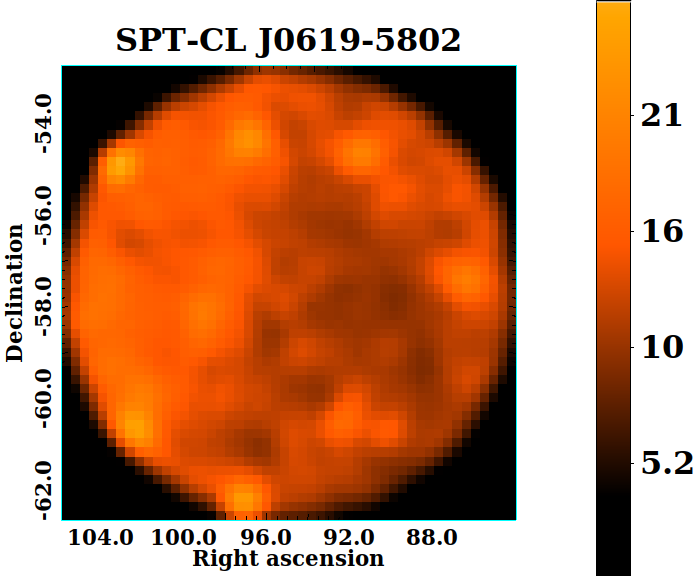}
\end{adjustbox}
\caption{Gaussian kernel smoothed counts map of the SPT-CL J0619-5802 cluster done in the same way as in Fig.~\ref{fig:countmap2012}.}
\label{fig:figure9}
\end{figure}

\begin{figure}
\centering
\begin{adjustbox}{right=1\columnwidth}
\includegraphics[width=1\columnwidth]{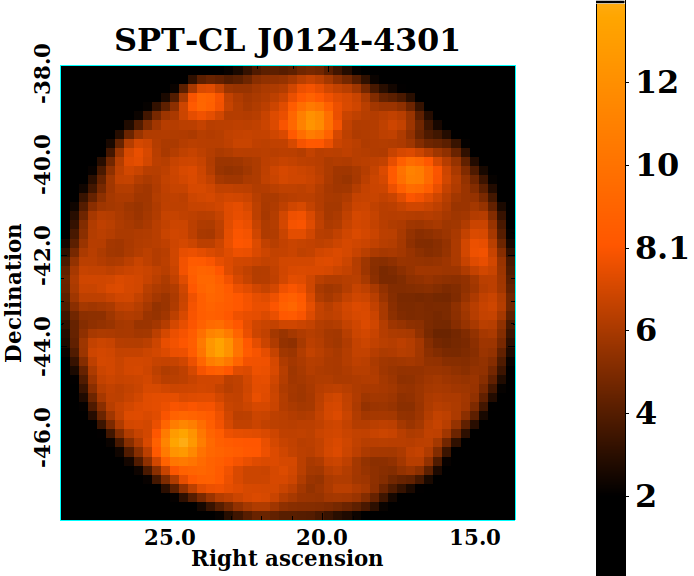}
\end{adjustbox}
\caption{Gaussian kernel smoothed counts map of the SPT-CL J0124-4301 cluster done in the same way as in Fig.~\ref{fig:countmap2012}}
\label{fig:figure10}
\end{figure}

\begin{figure}
\centering
\begin{adjustbox}{right=1\columnwidth}
\includegraphics[width=1\columnwidth]{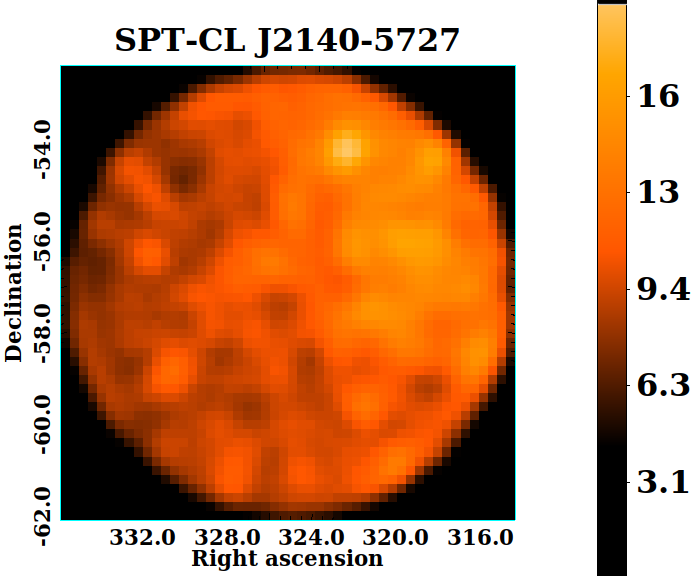}
\end{adjustbox}
\caption{Gaussian kernel smoothed counts map of the SPT-CL J2140-5727 cluster done in the same way as in Fig.~\ref{fig:countmap2012}.}
\label{fig:figure11}
\end{figure}

%\clearpage
%\clearpage
We now discuss the  remaining  six clusters with significance greater than $3\sigma$. These clusters include SPT-CL J2021-5257, SPT-CL J0217-5245, SPT-CL J0232-5257, SPT-CL J0619-5802, SPT-CL J0124-4301, and SPT-CL J2140-5727. The count maps after smoothening with a Gaussian kernel ($\sigma=1.5$) are depicted in Figures ~\ref{fig:figure6},~\ref{fig:figure7},~\ref{fig:figure8},~\ref{fig:figure9},~\ref{fig:figure10}, and ~\ref{fig:figure11}, respectively. We have shown their corresponding  TS maps after smoothening with a Gaussian kernel in Figures~\ref{fig:figure15}, ~\ref{fig:figure16}, ~\ref{fig:figure17}, ~\ref{fig:figure18}, ~\ref{fig:figure19}, and ~\ref{fig:figure20} respectively. 
For three of these clusters, SPT-CL J0217-5245, SPT-CL J0619-5802 and SPT-CL J2140-5727, we were able to find the counterparts SUMSS J021714-524528, SUMSS J061941-580217 and SUMSS J214034-572717, respectively in the SUMSS radio source catalogue within one arcminute,  with the integrated 36-cm flux density of these sources equal to 23.5 mJy, 33.2 mJy, and 9.0 mJy, respectively. These radio sources have also been classified as radio galaxies and could contribute to the observed emission for these clusters.
%For all six clusters with $TS>9$, we did not find coincidences within $0.2^{\circ}$ latest Fermi catalogue 4FGL-DR4 catalogue. 
%For all SPT clusters with $TS>9$, we calculated the upper limit energy emission in the energy range 1000 MeV to 300 GeV. The upper limits are found to be, for SPT-CL J2021-5257, it is $1.32 \times 10^{-9}$ ph cm$^{-2}$ s$^{-1}$; for SPT-CL J0217-5245,  we found it to be $1.07 \times 10^{-9}$ ph cm$^{-2}$ s$^{-1}$; similarly for SPT-CL J2140-5727, it is $0.53 \times 10^{-9}$ ph cm$^{-2}$ s$^{-1}$; and for SPT-CL J0232-5257, it is found to be $1.02 \times 10^{-9}$ ph cm$^{-2}$ s$^{-1}$; for SPT-CL J0124-4301, it is $0.63 \times 10^{-9}$ ph cm$^{-2}$ s$^{-1}$; and for SPT-CL J0619-5802, it is $0.82 \times 10^{-9}$ ph cm$^{-2}$ s$^{-1}$.

%\clearpage

\begin{figure}
\centering
\begin{adjustbox}{right=1\columnwidth}
\includegraphics[width=1\columnwidth]{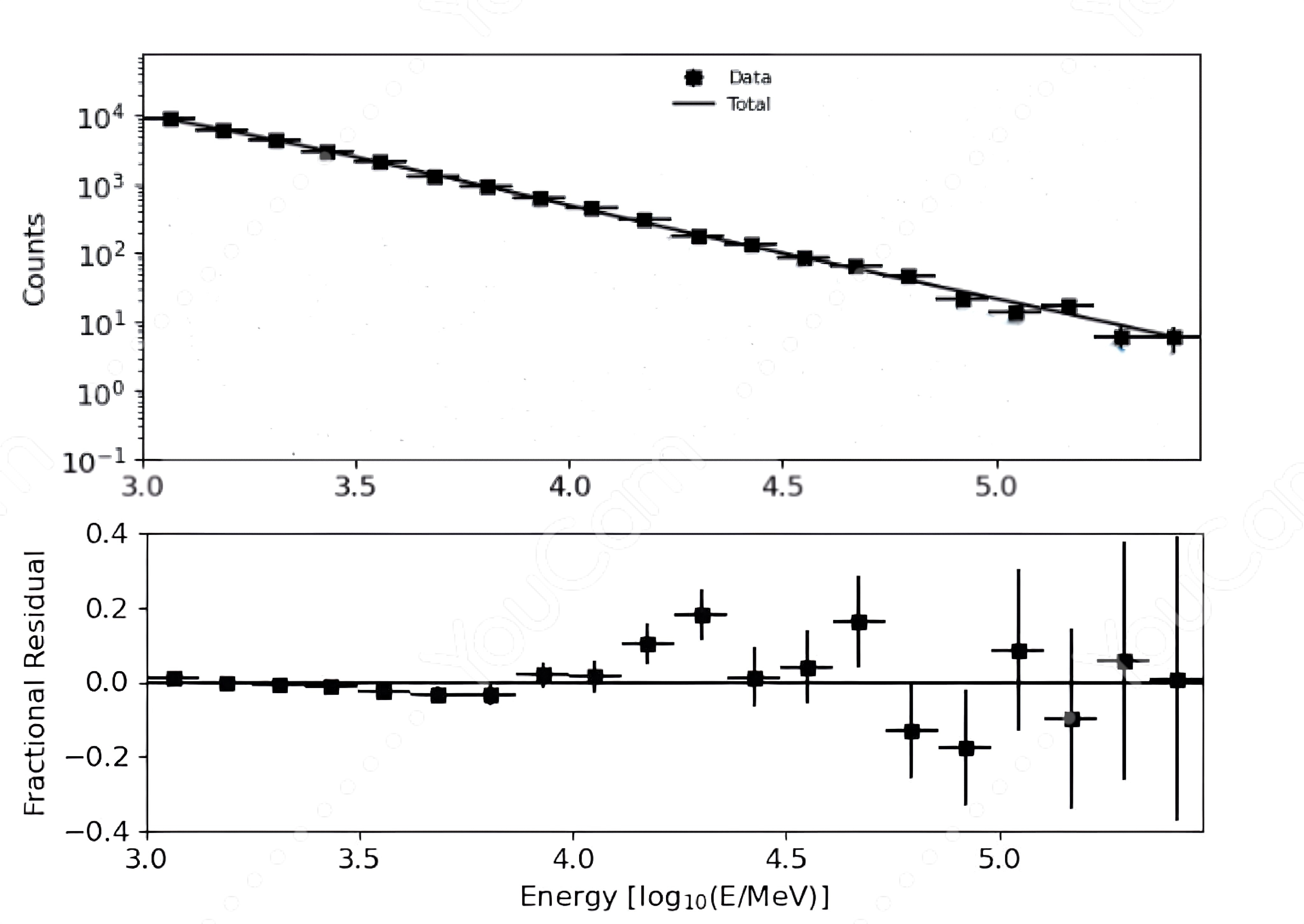}
\end{adjustbox}
\caption{Observed photons and residuals for SPT-CL J2021-5257 in the energy range 1 GeV to 300 GeV. This plot  uses the same specifications as in Fig.~\ref{fig:figure3}.} 
\label{fig:figure12}
\end{figure}

\begin{figure}
\centering
\begin{adjustbox}{right=1\columnwidth}
\includegraphics[width=1\columnwidth]{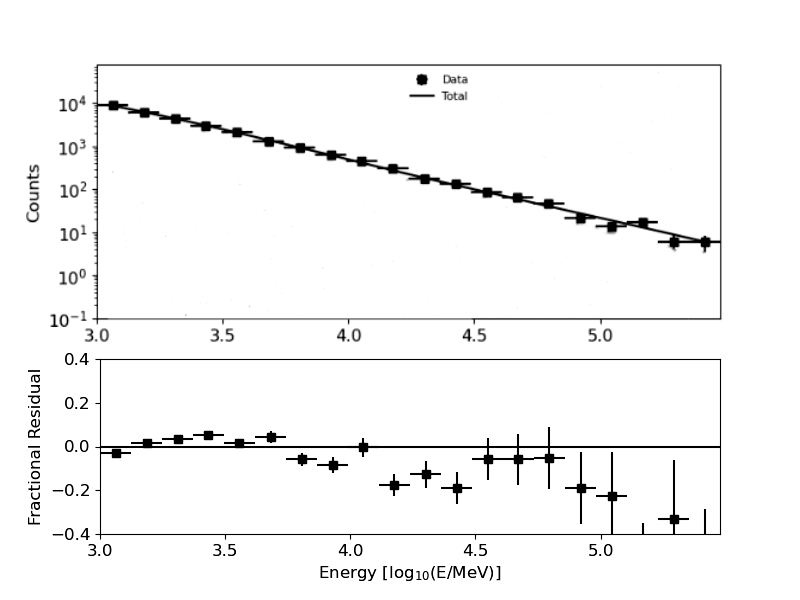}
\end{adjustbox}
\caption{ Observed photons and residuals for SPT-CL J0619-5802 in the energy range 1 GeV to 300 GeV. This plot uses  the same specifications as in Fig.~\ref{fig:figure3}.}
\label{fig:figure13}
\end{figure}

\begin{figure}[!t]
\centering
\begin{adjustbox}{right=1\columnwidth}
\includegraphics[width=1\columnwidth]{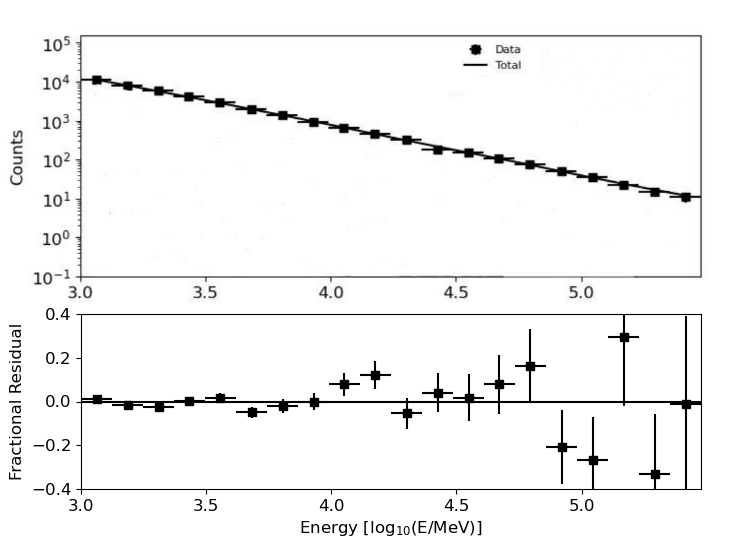}
\end{adjustbox}
\caption{Observed photons and residuals  for SPT-CL J0217-5245 in the energy range 1 GeV to 300 GeV. This plot uses  the same specifications as in Fig.~\ref{fig:figure3}.}
\label{fig:figure14}
\end{figure}

\begin{figure}
\centering
\begin{adjustbox}{right=1\columnwidth}
\includegraphics[width=1\columnwidth]{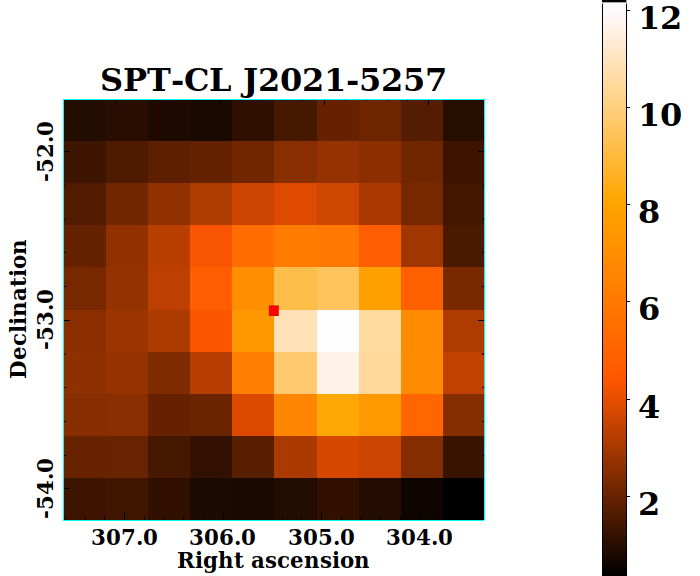}
\end{adjustbox}
\caption{Gaussian kernel smoothed TS map of the SPT-CL J2021-5257 cluster done in the same way as Fig.~\ref{fig:TS2012}.}
\label{fig:figure15}
\end{figure}

\begin{figure}
\centering
\begin{adjustbox}{right=1\columnwidth}
\includegraphics[width=1\columnwidth]{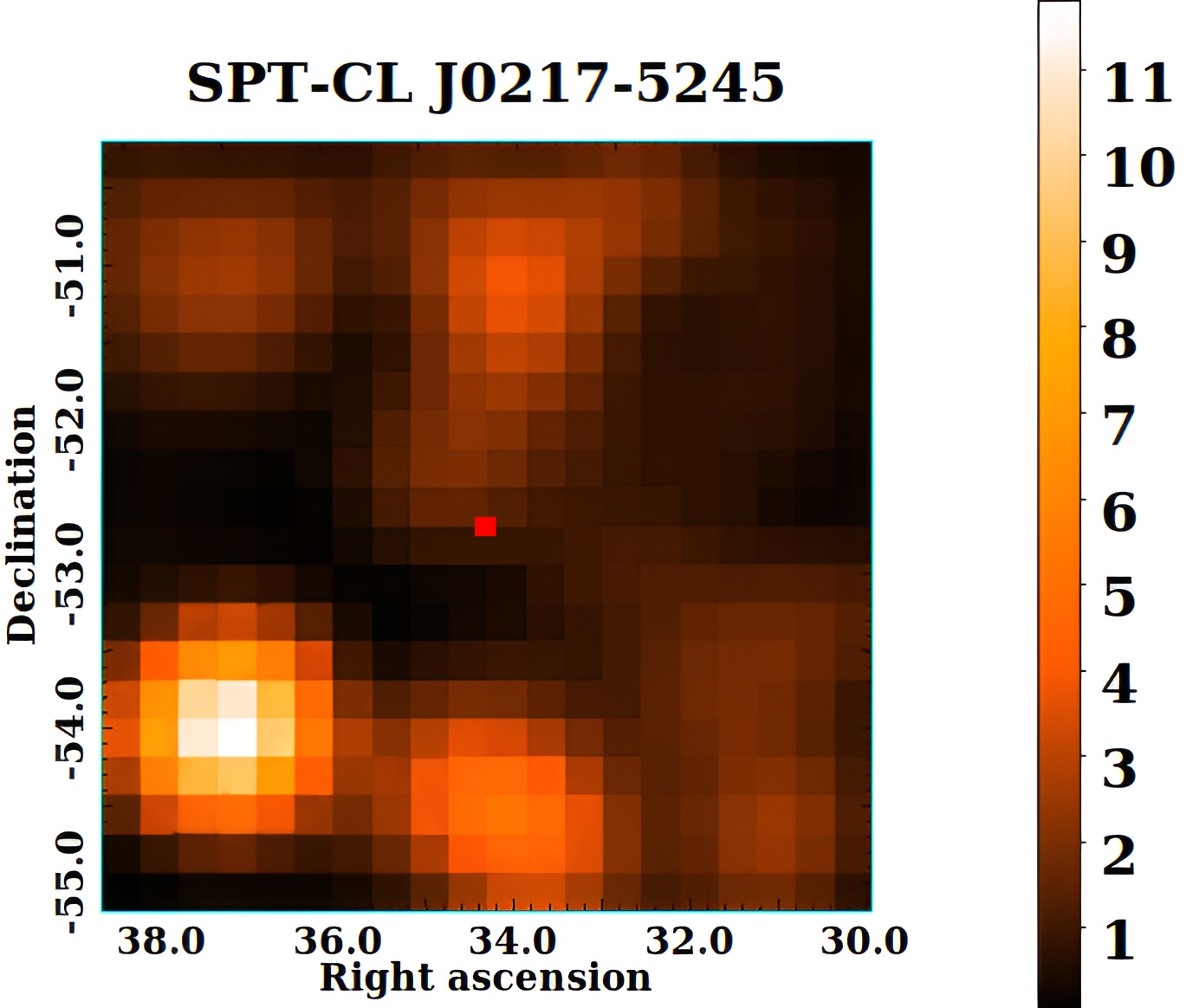}
\end{adjustbox}
\caption{Gaussian kernel smoothed TS map of the SPT-CL J0217-5245 cluster done in the same way as Fig.~\ref{fig:TS2012}.}
\label{fig:figure16}
\end{figure}

\begin{figure}
\centering
\begin{adjustbox}{right=1\columnwidth}
\includegraphics[width=1\columnwidth]{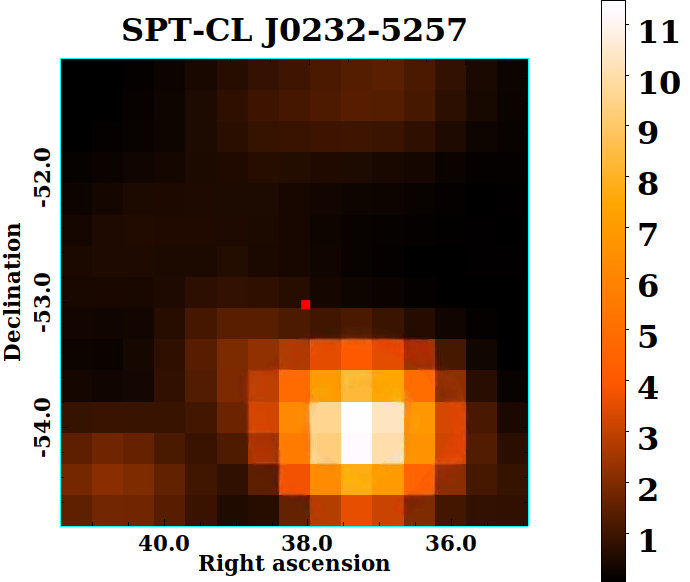}
\end{adjustbox}
\caption{Gaussian kernel smoothed TS map of the SPT-CL J0232-5257 cluster done in the same way as Fig.~\ref{fig:TS2012}.}
\label{fig:figure17}
\end{figure}

\begin{figure}
\centering
\begin{adjustbox}{right=1\columnwidth}
\includegraphics[width=1\columnwidth]{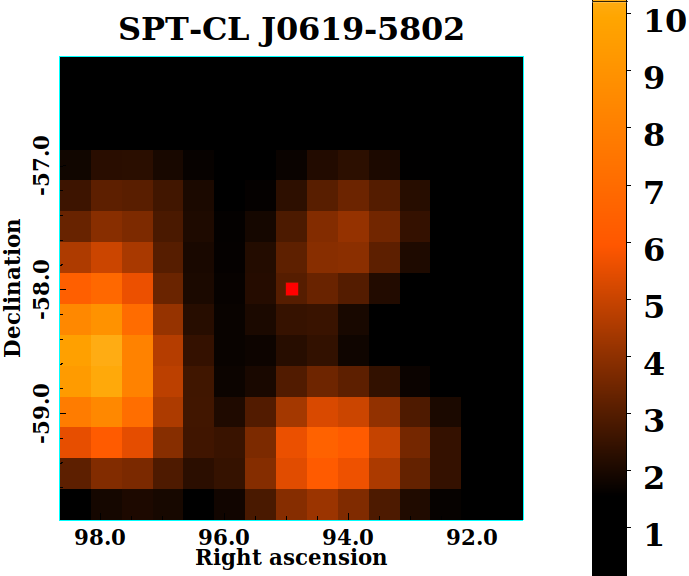}
\end{adjustbox}
\caption{Gaussian kernel smoothed TS map of the SPT-CL J0619-5802 cluster done in the same way as Fig.~\ref{fig:TS2012}.}
\label{fig:figure18}
\end{figure}

\begin{figure}
\centering
\begin{adjustbox}{right=1\columnwidth}
\includegraphics[width=1\columnwidth]{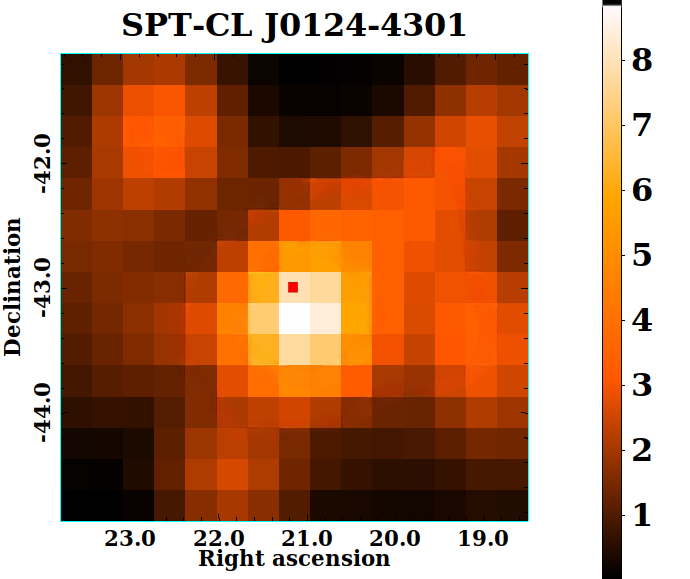}
\end{adjustbox}
\caption{Gaussian kernel smoothed TS map of the SPT-CL J0124-4301 cluster done in the same way as Fig.~\ref{fig:TS2012}.}
\label{fig:figure19}
\end{figure}

\begin{figure}
\centering
\begin{adjustbox}{right=1\columnwidth}
\includegraphics[width=1\columnwidth]{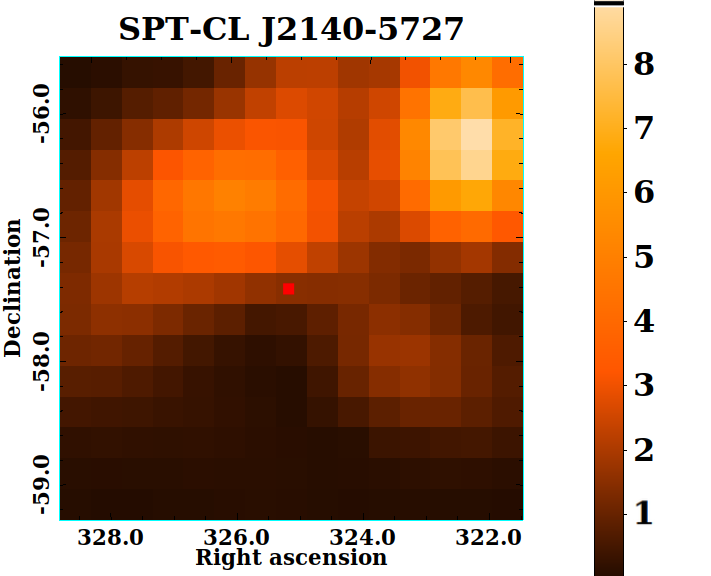}
\end{adjustbox}
\caption{Gaussian kernel smoothed TS map of the SPT-CL J2140-5727 cluster done in the same way as Fig.~\ref{fig:TS2012}.}
\label{fig:figure20}
\end{figure}
%\clearpage
\section{\label{sec:conclusions}Conclusions\protect}
In this work, we  have searched for gamma-ray emission with energies between 1-300 GeV from galaxy clusters, selected from the SPT-SZ 2500 sq. degree survey. For this purpose, we  used 15 years of  Fermi-LAT data and point-source templates for these searches. This analysis was done   using  300 SPT-SZ galaxy clusters, after sorting them in descending order based on their $M_{500}/z^{2}$ values. Among these clusters, we found statistically significant emission from SPT-CL J2012-5649 (Abell 3667) with  a detection significance of $6.1\sigma$. This signal is confined to $0.2 R_{200}$. The  total  photon flux is approximately  equal to  $1.3  \times 10^{-10}$ ph cm$^{-2}$ s$^{-1}$ with   the total energy flux  $\sim 1.3 \times 10^{-6}$ MeV cm$^{-2}$ s$^{-1}$. This cluster is a merging cluster for which radio relics have been detected using MeerKAT observations. The detection significance reduces to about $2.5\sigma$ if we use non-point source templates or search in a higher energy range from 10  to 300 GeV.  The SED for this cluster can be found in Fig.~\ref{fig:SEDplot}. The signal is observed  upto 10 GeV, while the spectral index is equal to $-3.61 \pm 0.33$. Above 10 GeV, we only obtain upper limits, which is consistent with the results from the MLE analysis in this energy range. Although prima-facie,  this constitutes only  the second galaxy cluster after Coma with detection significance $> 5\sigma$ at GeV energies, we note that  there are  six radio galaxies from the SUMSS catalogue within $0.2^{\circ}$ of this cluster, which is within  the Fermi-LAT PSF at this energy.  Therefore, we cannot definitely conclude that the gamma-ray emission detected  for SPT-CL J2012-5649 is coming from the ICM  as opposed to radio sources contributing to this emission.

We also found six other clusters  with significance between $3-5\sigma$. These clusters include SPT-CL J2021-5257, SPT-CL J0217-5245, SPT-CL J0232-5257, SPT-CL J1024-4301, SPT-CL J0619-5802, and SPT-CL J2140-5727. For three of these clusters (SPT-CL J0217-5245, SPT-CL J0619-5802 and SPT-CL J2140-5727) we again found three SUMSS radio galaxies  within one arcminute.  None of the remaining clusters show any evidence of gamma-ray emission. Two of these clusters (with null results) include the Bullet and El Gordo clusters.

In future works, we shall extend this analysis to all the SPT-SZ clusters (including those from SPTPol),  redo the search with other search templates, using the observed radio emission as well as  from dark matter annihilations.

%While we extended our analysis to 299 other SPT-SZ galaxy clusters, none exhibited detections exceeding 5$\sigma$. However, several noteworthy sources showed detections at around 3$\sigma$, contributing to our understanding of high-energy astrophysics. Furthermore, Abell 3667's unique characteristics, such as a merging cluster with shock waves and radio relics, make it a special object of study in galaxy cluster research. The collisions, shock waves, and galaxy mergers within Abell 3667 provide valuable insights into the formation and evolution of galaxy clusters, making it a highly significant target for astronomers. Our study advances our knowledge of gamma-ray emissions from galaxy clusters. Abell 3667 stands out as a remarkable object of study with its intriguing astrophysical features and high-significance gamma-ray emissions. These findings contribute to our broader understanding of the universe's largest structures.

\begin{acknowledgments}

 We  also appreciate the invaluable contributions of the Fermi-LAT team for making the Fermi-LAT data and analysis codes publicly available and answering all our queries. Without their state-of-the-art analysis technique, this research would not have been possible.   We acknowledge the National Supercomputing Mission (NSM) for providing computing resources of ‘PARAM SEVA’ at IIT, Hyderabad, which is implemented by C-DAC and supported by the Ministry of Electronics and Information Technology (MeitY) and Department of Science and Technology (DST), Government of India. This research has used SAOImageDS9, developed by Smithsonian   Astrophysical Observatory and we also thank them. We would also like to thank {\tt easyFermi} team \cite{deMenezes22} and {\tt Fermipy} team \cite{Wood17} for their software and support.
\end{acknowledgments}

%\clearpage
%\bibliographystyle{apsrev4-1} % Choose a bibliography style; here, we're using the APS style
\bibliography{references}
\newpage
\clearpage
\begin{center}
\begin{longtable*}{|c|c|c|c|c|c|}
\hline
\endfirsthead

\multicolumn{6}{c}%
{{\bfseries \tablename\ \thetable{} -- continued}} \\
\hline
\endhead

\hline
\endfoot

\endlastfoot

%Your table data starts here
\textbf{Cluster Name}  & \textbf{RA} & \textbf{Dec} & $\mathbf{M_{500}}$ & $\mathbf{z}$ & \textbf{TS Values}  \\
&  ($^{\circ}$) & ($^{\circ}$)  & $ (\times 10^{14} M_{\odot})/h $ & & \\   \hline \hline
\textbf{SPT-CL J2012-5649} & \textbf{303.11} & \textbf{-56.83} & \textbf{4.85}  & \textbf{0.06} & \textbf{37.2} \\
SPT-CL J0431-6126 & 67.84 & -61.44 & 4.84 & 0.06 & 3.0 \\
SPT-CL J2313-4243 & 348.50 & -42.73 & 4.52 & 0.06 & 3.0 \\
SPT-CL J2009-4518 & 302.45 & -45.31 & 4.20 & 0.06 & 2.0 \\
SPT-CL J2201-5956 & 330.47 & -59.94 & 9.27 & 0.10 & 0.6 \\
SPT-CL J0328-5541 & 52.17 & -55.70 & 5.47 & 0.08  & 5.0 \\
SPT-CL J2217-6509 & 334.49 & -65.15 & 5.34 & 0.09 & 2.0 \\
SPT-CL J2249-6426 & 342.43 & -64.43 & 4.83 & 0.09 & 3.0 \\
SPT-CL J0145-5301 & 26.26 & -53.03 & 5.61 & 0.11 & 3.0 \\
SPT-CL J0645-5413 & 101.37 & -54.22 & 10.05 & 0.16 & 1.0 \\
SPT-CL J0500-5116 & 75.24 & -51.27 & 4.46 & 0.11 & 3.0 \\
SPT-CL J2055-5456 & 313.99 & -54.94 & 5.64 & 0.14 & 2.0 \\
SPT-CL J0628-4143 & 97.20 & -41.73 & 8.46 & 0.18 & 3.0 \\
SPT-CL J0404-6510 & 61.05 & -65.18 & 3.85 & 0.12 & 4.0 \\
SPT-CL J0641-5001 & 100.46 & -50.02 & 3.92 & 0.12 & 7.0 \\
SPT-CL J0411-6340 & 62.86 & -63.68 & 4.85 & 0.14 & 3.0 \\
SPT-CL J0145-6033 & 26.30 & -60.56 & 7.17 & 0.18 & 0.5 \\
\textbf{SPT-CL J2021-5257} & \textbf{305.48} & \textbf{-52.95} & \textbf{4.24} & \textbf{0.14} & \textbf{12.0} \\
SPT-CL J0638-5358 & 99.70 & -53.97 & 11.29 & 0.23 & 2.0 \\
SPT-CL J2012-4130 & 303.00 & -41.50 & 4.66 & 0.15 & 1.5 \\
SPT-CL J0516-6312 & 79.09 & -63.21 & 4.00 & 0.14 & 8.0 \\
SPT-CL J2259-5617 & 344.00 & -56.29 & 4.71 & 0.15 & 3.0 \\
SPT-CL J0027-5015 & 6.82 & -50.25 & 4.24 & 0.15 & 2.5 \\
SPT-CL J0525-4715 & 81.46 & -47.26 & 7.23 & 0.19 & 3.0 \\
SPT-CL J2254-5805 & 343.59 & -58.09 & 4.47 & 0.15 & 3.0 \\
SPT-CL J0537-6504 & 84.35 & -65.07 & 7.02 & 0.20 & 8.5 \\
SPT-CL J0658-5556 & 104.63 & -55.95 & 15.38 & 0.30 & 2.1 \\
SPT-CL J0510-4519 & 77.58 & -45.33 & 6.94 & 0.20 & 3.5 \\
SPT-CL J0637-4829 & 99.35 & -48.49 & 6.85 & 0.20 & 6.3 \\
SPT-CL J2134-4238 & 323.50 & -42.64 & 6.23 & 0.19 & 6.1 \\
SPT-CL J0051-4834 & 12.79 & -48.58 & 5.69 & 0.19 & 2.6 \\
SPT-CL J0317-4849 & 49.45 & -48.83 & 4.05 & 0.16 & 5.1 \\
SPT-CL J2023-5535 & 305.84 & -55.59 & 7.77 & 0.23 & 4.5 \\
SPT-CL J0232-4421 & 38.07 & -44.35 & 11.29 & 0.28 & 6.3 \\
SPT-CL J0216-4816 & 34.07 & -48.28 & 4.07 & 0.17 & 4.7 \\
SPT-CL J2025-5117 & 306.48 & -51.29 & 6.61 & 0.22 & 8.4 \\
SPT-CL J0010-5112 & 2.74 & -51.21 & 3.91 & 0.17 & 2.5 \\
SPT-CL J2254-6314 & 343.51 & -63.25 & 5.94 & 0.21 & 5.8 \\
SPT-CL J2248-4431 & 342.19 & -44.53 & 15.71 & 0.35 & 2.2 \\
SPT-CL J0013-4621 & 3.47 & -46.36 & 3.91 & 0.18 & 5.4 \\
SPT-CL J2020-4646 & 305.19 & -46.77 & 4.21 & 0.19 & 2.0 \\
SPT-CL J0504-4929 & 76.01 & -49.49 & 4.53 & 0.20 & 3.0 \\
SPT-CL J0225-4155 & 36.48 & -41.92 & 5.20 & 0.22 & 2.0 \\
SPT-CL J0458-5741 & 74.60 & -57.70 & 3.88 & 0.19 & 6.3 \\
SPT-CL J2241-4236 & 340.47 & -42.60 & 4.25 & 0.20 & 2.5 \\
SPT-CL J0108-4341 & 17.13 & -43.69 & 3.83 & 0.19 & 3.6 \\
SPT-CL J0124-5937 & 21.20 & -59.63 & 4.65 & 0.21 & 5.0 \\
SPT-CL J0256-4736 & 44.24 & -47.61 & 5.30 & 0.23 & 7.5 \\
SPT-CL J0653-5744 & 103.33 & -57.75 & 5.71 & 0.24 & 3.7 \\
SPT-CL J2211-4833 & 332.83 & -48.56 & 5.50 & 0.24 & 2.5 \\
SPT-CL J0118-5638 & 19.54 & -56.63 & 4.15 & 0.21 & 2.6 \\
SPT-CL J0651-4037 & 102.82 & -40.63 & 5.15 & 0.24 & 5.7 \\
SPT-CL J2121-6335 & 320.43 & -63.58 & 4.21 & 0.22 & 3.2 \\
SPT-CL J0001-6258 & 0.40 & -62.98 & 3.89 & 0.21 & 2.3 \\
SPT-CL J0235-5121 & 38.95 & -51.35 & 6.56 & 0.28 & 1.7 \\
SPT-CL J2005-5635 & 301.34 & -56.59 & 3.71 & 0.21 & 4.7 \\
SPT-CL J2254-4620 & 343.59 & -46.34 & 5.93 & 0.27 & 3.7 \\
SPT-CL J0549-6205 & 87.33 & -62.09 & 11.65 & 0.37 & 3.4 \\
SPT-CL J0516-5430 & 79.15 & -54.51 & 7.12 & 0.29 & 4.6 \\
SPT-CL J2031-4037 & 307.97 & -40.62 & 9.4 & 0.34 & 3.0 \\
SPT-CL J0225-4327 & 36.30 & -43.46 & 4.25 & 0.23 & 3.8 \\
SPT-CL J0040-4407 & 10.20 & -44.13 & 9.8 & 0.35 & 2.2 \\
SPT-CL J0620-4715 & 95.10 & -47.26 & 4.22 & 0.23 & 1.9 \\
SPT-CL J2019-5642 & 304.77 & -56.71 & 4.05 & 0.23 & 5.3 \\
SPT-CL J0249-5658 & 42.41 & -56.98 & 4.19 & 0.23 & 2.1 \\
SPT-CL J0603-4714 & 90.99 & -47.24 & 5.63 & 0.27 & 2.2 \\
SPT-CL J0601-4122 & 90.50 & -41.37 & 3.92 & 0.23 & 4.4 \\
SPT-CL J2138-6008 & 324.51 & -60.13 & 7.27 & 0.32 & 1.0 \\
SPT-CL J2032-5627 & 308.08 & -56.46 & 5.74 & 0.28 & 1.7 \\
SPT-CL J2223-5015 & 335.81 & -50.27 & 4.06 & 0.24 & 2.9 \\
SPT-CL J0412-5106 & 63.23 & -51.11 & 4.00 & 0.24 & 3.9 \\
SPT-CL J2120-4016 & 320.14 & -40.27 & 4.20 & 0.25 & 2.7 \\
SPT-CL J2219-5708 & 334.96 & -57.14 & 6.30 & 0.31 & 4.1 \\
SPT-CL J2300-5331 & 345.18 & -53.52 & 4.47 & 0.26 & 6.0 \\
SPT-CL J2027-4240 & 306.93 & -42.67 & 3.95 & 0.25 & 2.4 \\
SPT-CL J0555-6406 & 88.87 & -64.10 & 7.64 & 0.34 & 6.0 \\
SPT-CL J0311-6354 & 47.83 & -63.91 & 5.14 & 0.28 & 3.1 \\
SPT-CL J0129-6432 & 22.43 & -64.54 & 6.73 & 0.32 & 3.9 \\
SPT-CL J0505-6145 & 76.40 & -61.75 & 5.15 & 0.29 & 8.2 \\
SPT-CL J0133-6434 & 23.41 & -64.57 & 6.26 & 0.32 & 3.4 \\
SPT-CL J2325-4111 & 351.30 & -41.20 & 7.52 & 0.36 & 1.7 \\
SPT-CL J0438-5419 & 69.57 & -54.32 & 10.28 & 0.42 & 4.9 \\
SPT-CL J0405-4916 & 61.49 & -49.27 & 5.08 & 0.30 & 2.1 \\
SPT-CL J2344-4224 & 356.15 & -42.41 & 3.74 & 0.26 & 5.6 \\
SPT-CL J0150-4511 & 27.65 & -45.19 & 5.28 & 0.31 & 1.9 \\
SPT-CL J0041-4428 & 10.25 & -44.48 & 6.05 & 0.33 & 0 \\
SPT-CL J0214-4638 & 33.70 & -46.65 & 4.90 & 0.30 & 5.4 \\
SPT-CL J0143-4452 & 25.89 & -44.87 & 4.00 & 0.27 & 2.9 \\
SPT-CL J0655-5541 & 103.91 & -55.69 & 4.44 & 0.29 & 1.9 \\
SPT-CL J2101-5542 & 315.31 & -55.70 & 3.82 & 0.27 & 3.0 \\
SPT-CL J0440-4657 & 70.23 & -46.97 & 5.05 & 0.31 & 2.0 \\
SPT-CL J0106-5943 & 16.62 & -59.72 & 6.40 & 0.35 & 4.3 \\
SPT-CL J2130-6458 & 322.73 & -64.98 & 5.23 & 0.32 & 4.2 \\
SPT-CL J0439-4600 & 69.81 & -46.01 & 5.58 & 0.33 & 4.4 \\
SPT-CL J2011-5725 & 302.85 & -57.42 & 3.99 & 0.28 & 5.6 \\
SPT-CL J0348-4515 & 57.07 & -45.25 & 6.33 & 0.36 & 3.2 \\
SPT-CL J0304-4921 & 46.06 & -49.36 & 7.53 & 0.39 & 5.2 \\
SPT-CL J0455-4159 & 73.99 & -41.99 & 4.01 & 0.29 & 6.1 \\
SPT-CL J2223-5227 & 335.87 & -52.47 & 4.00 & 0.29 & 1.3 \\
SPT-CL J2115-4659 & 318.80 & -46.99 & 4.25 & 0.30 & 4.2 \\
SPT-CL J2016-4954 & 304.01 & -49.91 & 3.91 & 0.29 & 2.6 \\
SPT-CL J0522-4818 & 80.57 & -48.30 & 4.08 & 0.30 & 6.3 \\
SPT-CL J0151-5654 & 27.79 & -56.91 & 3.86 & 0.29 & 3.1 \\
SPT-CL J0234-5831 & 38.68 & -58.52 & 7.93 & 0.41 & 4.5 \\
SPT-CL J0114-4123 & 18.68 & -41.39 & 7.03 & 0.39 & 4.3 \\
SPT-CL J0022-4144 & 5.55 & -41.74 & 4.05 & 0.30 & 3.2 \\
SPT-CL J0411-4819 & 62.82 & -48.32 & 8.04 & 0.42 & 3.0 \\
SPT-CL J0551-4339 & 87.88 & -43.66 & 4.82 & 0.33 & 3.7 \\
SPT-CL J2355-5055 & 358.95 & -50.93 & 4.49 & 0.32 & 3.9 \\
SPT-CL J0509-6118 & 77.47 & -61.31 & 6.65 & 0.39 & 4.9 \\
SPT-CL J0636-4942 & 99.17 & -49.70 & 5.40 & 0.35 & 7.5 \\
SPT-CL J0110-4445 & 17.59 & -44.76 & 5.57 & 0.36 & 3.0 \\
SPT-CL J0013-4906 & 3.33 & -49.12 & 7.12 & 0.41 & 3.9 \\
SPT-CL J0001-4842 & 0.28 & -48.71 & 4.48 & 0.33 & 3.7 \\
\textbf{SPT-CL J0217-5245} & \textbf{34.30} & \textbf{-52.76} & \textbf{4.82} & \textbf{0.34} & \textbf{11.9} \\
SPT-CL J0254-5857 & 43.57 & -58.95 & 7.72 & 0.44 & 3.1 \\
SPT-CL J2330-4502 & 352.57 & -45.03 & 4.06 & 0.32 & 4.1 \\
SPT-CL J0304-4401 & 46.07 & -44.03 & 8.37 & 0.46 & 2.8 \\
SPT-CL J0236-4938 & 39.25 & -49.64 & 4.42 & 0.33 & 2.1 \\
SPT-CL J0600-4353 & 90.06 & -43.88 & 5.08 & 0.36 & 3.8 \\
SPT-CL J0052-5657 & 13.16 & -56.96 & 3.92 & 0.32 & 2.5 \\
SPT-CL J0650-4503 & 102.68 & -45.06 & 6.22 & 0.40 & 6.6 \\
SPT-CL J2059-5018 & 314.93 & -50.30 & 4.05 & 0.33 & 4.8 \\
SPT-CL J2131-4019 & 322.77 & -40.32 & 7.49 & 0.45 & 2.8 \\
SPT-CL J0416-6359 & 64.16 & -63.99 & 4.47 & 0.35 & 4.7 \\
SPT-CL J0424-4406 & 66.00 & -44.11 & 4.73 & 0.36 & 4.6 \\
SPT-CL J0240-5946 & 40.16 & -59.77 & 5.79 & 0.40 & 8.4 \\
SPT-CL J0144-4807 & 26.18 & -48.12 & 3.84 & 0.33 & 2.7 \\
SPT-CL J0405-4648 & 61.29 & -46.81 & 4.70 & 0.36 & 3.7 \\
SPT-CL J2358-6129 & 359.71 & -61.49 & 4.45 & 0.36 & 4.4 \\
SPT-CL J2135-5726 & 323.92 & -57.44 & 6.31 & 0.43 & 7.6 \\
SPT-CL J0330-5228 & 52.73 & -52.47 & 6.75 & 0.44 & 3.6 \\
SPT-CL J0412-5743 & 63.02 & -57.72 & 3.98 & 0.34 & 4.3 \\
SPT-CL J2332-5358 & 353.11 & -53.97 & 5.55 & 0.40 & 2.3 \\
SPT-CL J0012-5352 & 3.06 & -53.87 & 3.91 & 0.34 & 4.3 \\
SPT-CL J2205-5927 & 331.27 & -59.46 & 4.35 & 0.37 & 2.8 \\
SPT-CL J2344-4243 & 356.18 & -42.72 & 11.38 & 0.60 & 5.8 \\
SPT-CL J0402-4611 & 60.58 & -46.19 & 4.07 & 0.36 & 3.5 \\
SPT-CL J0052-4551 & 13.19 & -45.86 & 4.06 & 0.36 & 3.5 \\
SPT-CL J0243-4833 & 40.91 & -48.56 & 7.74 & 0.50 & 2.2 \\
SPT-CL J2022-6323 & 305.53 & -63.40 & 4.56 & 0.38 & 4.7 \\
SPT-CL J2327-5137 & 351.78 & -51.62 & 3.55 & 0.34 & 3.7 \\
SPT-CL J2316-5453 & 349.21 & -54.90 & 4.24 & 0.37 & 5.1 \\
SPT-CL J0344-5518 & 56.21 & -55.30 & 3.89 & 0.36 & 2.6 \\
SPT-CL J2145-5644 & 326.47 & -56.75 & 6.97 & 0.48 & 5.2 \\
SPT-CL J2233-5339 & 338.33 & -53.65 & 5.82 & 0.44 & 2.8 \\
SPT-CL J2159-6244 & 329.99 & -62.74 & 4.54 & 0.39 & 3.6 \\
SPT-CL J0445-4230 & 71.28 & -42.51 & 5.05 & 0.41 & 3.5 \\
SPT-CL J0551-5709 & 87.90 & -57.16 & 5.28 & 0.42 & 1.2 \\
SPT-CL J2124-6124 & 321.15 & -61.41 & 5.47 & 0.43 & 4.8 \\
SPT-CL J2351-5452 & 357.90 & -54.88 & 4.25 & 0.38 & 1.9 \\
SPT-CL J2206-4057 & 331.62 & -40.95 & 3.70 & 0.36 & 5.4 \\
SPT-CL J0252-4824 & 43.19 & -48.41 & 5.04 & 0.42 & 7.6 \\
SPT-CL J0429-4355 & 67.31 & -43.93 & 3.83 & 0.37 & 5.3 \\
SPT-CL J2259-5431 & 344.98 & -54.53 & 4.23 & 0.39 & 3.9 \\
SPT-CL J0025-5034 & 6.37 & -50.57 & 3.85 & 0.37 & 8.3 \\
SPT-CL J0447-5055 & 71.84 & -50.92 & 4.29 & 0.40 & 2.8 \\
SPT-CL J0019-4051 & 4.76 & -40.86 & 6.18 & 0.48 & 3.2 \\
SPT-CL J2136-4704 & 324.12 & -47.08 & 4.78 & 0.43 & 1.7 \\
SPT-CL J0354-5904 & 58.56 & -59.07 & 4.57 & 0.42 & 3.8 \\
SPT-CL J0333-5842 & 53.32 & -58.70 & 3.59 & 0.37 & 1.8 \\
SPT-CL J0244-4857 & 41.03 & -48.96 & 4.41 & 0.41 & 2.4 \\
SPT-CL J0342-4028 & 55.56 & -40.48 & 4.92 & 0.44 & 5.3 \\
SPT-CL J2030-5638 & 307.70 & -56.64 & 3.97 & 0.39 & 2.9 \\
SPT-CL J0509-5342 & 77.34 & -53.71 & 5.35 & 0.46 & 8.5 \\
SPT-CL J2319-4716 & 349.98 & -47.28 & 4.54 & 0.43 & 2.9 \\
SPT-CL J0611-5938 & 92.81 & -59.64 & 3.77 & 0.39 & 6.3 \\
SPT-CL J0054-4046 & 13.59 & -40.78 & 4.11 & 0.40 & 5.7 \\
SPT-CL J0505-4204 & 76.37 & -42.08 & 3.89 & 0.40 & 5.6 \\
SPT-CL J0655-5234 & 103.96 & -52.57 & 5.42 & 0.47 & 2.8 \\
SPT-CL J0334-4659 & 53.55 & -46.99 & 5.76 & 0.49 & 3.3 \\
SPT-CL J2016-4517 & 304.00 & -45.30 & 3.86 & 0.40 & 5.0 \\
SPT-CL J2342-4714 & 355.75 & -47.24 & 3.81 & 0.40 & 5.8 \\
SPT-CL J0047-4506 & 11.82 & -45.11 & 5.24 & 0.47 & 4.2 \\
SPT-CL J0439-5330 & 69.93 & -53.50 & 4.06 & 0.41 & 6.5 \\
SPT-CL J0647-5828 & 101.98 & -58.48 & 4.64 & 0.44 & 6.1 \\
SPT-CL J0341-5027 & 55.28 & -50.46 & 3.59 & 0.39 & 2.2 \\
SPT-CL J0254-6051 & 43.60 & -60.86 & 4.60 & 0.44 & 5.7 \\
\textbf{SPT-CL J2140-5727} & \textbf{325.14} & \textbf{-57.46} & \textbf{3.87} & \textbf{0.40} & \textbf{9.0} \\
SPT-CL J0452-4806 & 73.00 & -48.11 & 3.47 & 0.38 & 3.1 \\
SPT-CL J0259-4556 & 44.90 & -45.94 & 4.28 & 0.43 & 4.6 \\
SPT-CL J0626-4446 & 96.74 & -44.77 & 4.50 & 0.44 & 3.9 \\
SPT-CL J0237-4151 & 39.42 & -41.86 & 3.78 & 0.41 & 2.2 \\
SPT-CL J2035-5251 & 308.80 & -52.85 & 6.38 & 0.53 & 4.6 \\
SPT-CL J0014-4036 & 3.74 & -40.60 & 6.14 & 0.52 & 3.0 \\
SPT-CL J0257-5842 & 44.39 & -58.71 & 3.91 & 0.42 & 2.3 \\
SPT-CL J0638-4243 & 99.57 & -42.72 & 3.71 & 0.41 & 6.9 \\
SPT-CL J0417-4748 & 64.35 & -47.81 & 7.41 & 0.58 & 5.0 \\
SPT-CL J0655-4429 & 103.76 & -44.48 & 3.72 & 0.41 & 6.1 \\
SPT-CL J0543-6219 & 85.76 & -62.32 & 5.53 & 0.51 & 4.3 \\
SPT-CL J2335-4544 & 353.79 & -45.74 & 6.38 & 0.55 & 5.3 \\
SPT-CL J2306-6505 & 346.73 & -65.09 & 5.96 & 0.53 & 6.6 \\
SPT-CL J0611-4724 & 92.92 & -47.41 & 4.23 & 0.45 & 3.9 \\
SPT-CL J2112-4434 & 318.21 & -44.58 & 5.86 & 0.53 & 6.6 \\
SPT-CL J0038-5244 & 9.72 & -52.74 & 3.83 & 0.43 & 7.9 \\
SPT-CL J2111-5339 & 317.92 & -53.65 & 4.20 & 0.44 & 3.2 \\
SPT-CL J0351-5636 & 57.93 & -56.61 & 3.48 & 0.41 & 3.2 \\
SPT-CL J0546-4752 & 86.55 & -47.88 & 3.94 & 0.43 & 4.7 \\
SPT-CL J2050-4213 & 312.57 & -42.22 & 5.14 & 0.50 & 5.2 \\
SPT-CL J0200-4852 & 30.14 & -48.88 & 5.11 & 0.50 & 2.6 \\
SPT-CL J0216-4830 & 34.07 & -48.51 & 4.52 & 0.47 & 5.5 \\
SPT-CL J0346-5439 & 56.72 & -54.65 & 5.72 & 0.53 & 3.2 \\
SPT-CL J2259-3952 & 344.81 & -39.87 & 5.84 & 0.54 & 2.9 \\
SPT-CL J0257-4817 & 44.45 & -48.30 & 3.89 & 0.44 & 7.4 \\
SPT-CL J0451-4952 & 72.97 & -49.88 & 3.63 & 0.42 & 3.6 \\
SPT-CL J0532-5450 & 83.03 & -54.84 & 3.72 & 0.43 & 4.5 \\
SPT-CL J0317-5935 & 49.32 & -59.59 & 4.41 & 0.47 & 1.9 \\
SPT-CL J0257-5732 & 44.35 & -57.54 & 3.73 & 0.43 & 2.1 \\
SPT-CL J2235-4416 & 338.86 & -44.27 & 4.03 & 0.45 & 4.5 \\
SPT-CL J0517-6311 & 79.41 & -63.20 & 3.90 & 0.45 & 6.7 \\
SPT-CL J0403-5719 & 60.97 & -57.32 & 4.18 & 0.47 & 4.3 \\
SPT-CL J0456-6141 & 74.15 & -61.68 & 3.60 & 0.44 & 4.2 \\
SPT-CL J0314-6130 & 48.61 & -61.51 & 3.51 & 0.43 & 1.7 \\
SPT-CL J0508-6149 & 77.16 & -61.82 & 3.73 & 0.45 & 7.2 \\
SPT-CL J0641-5950 & 100.38 & -59.85 & 5.01 & 0.52 & 6.8 \\
SPT-CL J0417-4427 & 64.41 & -44.46 & 5.81 & 0.56 & 5.8 \\
SPT-CL J0307-5042 & 46.95 & -50.70 & 5.55 & 0.55 & 3.9 \\
\textbf{SPT-CL J0232-5257} & \textbf{38.19} & \textbf{-52.96} & \textbf{5.63} & \textbf{0.56} & \textbf{11.5} \\
\textbf{SPT-CL J0124-4301} & \textbf{21.15} & \textbf{-43.02} & \textbf{3.92} & \textbf{0.47} & \textbf{9.0} \\
SPT-CL J2132-4349 & 323.17 & -43.83 & 4.82 & 0.52 & 6.8 \\
SPT-CL J2131-5003 & 322.97 & -50.06 & 3.81 & 0.46 & 3.9 \\
SPT-CL J0102-4915 & 15.73 & -49.26 & 13.52 & 0.87 & 4.7 \\
SPT-CL J0113-6105 & 18.40 & -61.09 & 3.63 & 0.45 & 3.0 \\
SPT-CL J0304-4748 & 46.15 & -47.81 & 4.56 & 0.51 & 4.0 \\
SPT-CL J2331-5051 & 352.96 & -50.86 & 5.81 & 0.58 & 3.5 \\
SPT-CL J0544-3950 & 86.25 & -39.84 & 4.78 & 0.52 & 3.2 \\
SPT-CL J0048-4548 & 12.25 & -45.80 & 3.79 & 0.47 & 3.8 \\
SPT-CL J0007-4706 & 1.75 & -47.11 & 3.64 & 0.46 & 6.4 \\
SPT-CL J0306-4749 & 46.75 & -47.82 & 3.64 & 0.46 & 4.2 \\
SPT-CL J0337-6300 & 54.47 & -63.01 & 3.76 & 0.46 & 2.0 \\
SPT-CL J0111-5424 & 17.77 & -54.41 & 4.06 & 0.48 & 3.7 \\
SPT-CL J0456-5116 & 74.17 & -51.28 & 5.39 & 0.56 & 2.7 \\
SPT-CL J0257-6050 & 44.34 & -60.85 & 3.54 & 0.46 & 3.9 \\
SPT-CL J0659-5300 & 104.77 & -53.01 & 3.89 & 0.48 & 0 \\
SPT-CL J2145-4348 & 326.36 & -43.80 & 4.03 & 0.49 & 0 \\
SPT-CL J0253-6046 & 43.46 & -60.77 & 3.64 & 0.46 & 0 \\
SPT-CL J2302-4435 & 345.58 & -44.58 & 3.61 & 0.47 & 0 \\
SPT-CL J2245-6206 & 341.26 & -62.11 & 5.67 & 0.59 & 0 \\
SPT-CL J0337-4928 & 54.45  & -49.47 & 3.94 & 0.49 & 0 \\
SPT-CL J2232-5959 & 338.15 & -59.99 & 5.81 & 0.60 & 5.3 \\
SPT-CL J2040-5342 & 310.22 & -53.71 & 4.44 & 0.52 & 4.3 \\
SPT-CL J0559-5249 & 89.92 & -52.83 & 5.97 & 0.61 & 4.1 \\
SPT-CL J0307-6225 & 46.83 & -62.43 & 5.36 & 0.58 & 3.8 \\
SPT-CL J0025-4133 & 6.49 & -41.55 & 4.64 & 0.54 & 3.3 \\
SPT-CL J0109-4045 & 17.47 & -40.76 & 3.61 & 0.48 & 2.5 \\
SPT-CL J0238-4904 & 39.70 & -49.07 & 4.18 & 0.52 & 1.9 \\
SPT-CL J0221-4446 & 35.42 & -44.78 & 3.85 & 0.50 & 5.7 \\
SPT-CL J2051-6256 & 312.80 & -62.93 & 3.61 & 0.48 & 3.8 \\
SPT-CL J0218-4315 & 34.58 & -43.26 & 6.00 & 0.63 & 3.8 \\
SPT-CL J0106-5355 & 16.57 & -53.92 & 3.78 & 0.50 & 1.8 \\
SPT-CL J0219-4934 & 34.81 & -49.58 & 4.52 & 0.55 & 3.5 \\
SPT-CL J0111-5518 & 17.84 & -55.31 & 3.54 & 0.49 & 3.2 \\
SPT-CL J0351-4109 & 57.75 & -41.16 & 5.69 & 0.61 & 4.6 \\
SPT-CL J2017-6258 & 304.48 & -62.98 & 4.27 & 0.54 & 5.1 \\
SPT-CL J2148-6116 & 327.18 & -61.28 & 4.82 & 0.57 & 2.5 \\
\textbf{SPT-CL J0619-5802} & \textbf{94.92} & \textbf{-58.04} & \textbf{4.51} & \textbf{0.55} & \textbf{10.3} \\
SPT-CL J0135-5902 & 23.79 & -59.04 & 3.79 & 0.51 & 6.1 \\
SPT-CL J0041-5107 & 10.29 & -51.13 & 3.62 & 0.50 & 4.0 \\
SPT-CL J0342-5354 & 55.52 & -53.91 & 3.90 & 0.51 & 3.8 \\
SPT-CL J0135-5904 & 23.97 & -59.08 & 3.52 & 0.49 & 5.2 \\
SPT-CL J0343-5518 & 55.76 & -55.30 & 4.19 & 0.54 & 3.2 \\
SPT-CL J0033-6326 & 8.48 & -63.44 & 5.08 & 0.60 & 4.8 \\
SPT-CL J0212-4657 & 33.10 & -46.95 & 6.06 & 0.65 & 5.5 \\
SPT-CL J0410-6343 & 62.52 & -63.73 & 3.95 & 0.53 & 3.8 \\
SPT-CL J0336-4005 & 54.16 & -40.10 & 3.76 & 0.52 & 5.6 \\
SPT-CL J0217-4310 & 34.41 & -43.18 & 4.55 & 0.57 & 4.0 \\
SPT-CL J2218-4519 & 334.75 & -45.32 & 5.60 & 0.64 & 4.8 \\
SPT-CL J0429-5233 & 67.43 & -52.56 & 3.39 & 0.50 & 3.1 \\
SPT-CL J2337-5942 & 354.35 & -59.70 & 8.29 & 0.77 & 5.2 \\
SPT-CL J2155-6048 & 328.98 & -60.81 & 3.95 & 0.54 & 3.0 \\
SPT-CL J2350-5301 & 357.73 & -53.02 & 3.95 & 0.54 & 2.6 \\
SPT-CL J0542-4100 & 85.72 & -41.00 & 5.48 & 0.64 & 3.4 \\
SPT-CL J2222-4834 & 335.71 & -48.57 & 5.69 & 0.65 & 3.6 \\
SPT-CL J0522-5026 & 80.52 & -50.44 & 3.60 & 0.52 & 5.5 \\
SPT-CL J0309-4958 & 47.26 & -49.97 & 4.07 & 0.55 & 0 \\
SPT-CL J0202-5401 & 30.58 & -54.02 & 4.07 & 0.55 & 0 \\
SPT-CL J0426-5455 & 66.52 & -54.92 & 5.43 & 0.64 & 3.6 \\
SPT-CL J0142-5032 & 25.55 & -50.54 & 6.05 & 0.68 & 5.3 \\
SPT-CL J0402-6130 & 60.71 & -61.50 & 3.51 & 0.52 & 2.9 \\
SPT-CL J0011-4614 & 2.98 & -46.24 & 3.88 & 0.54 & 6.2 \\
SPT-CL J2155-5224 & 328.89 & -52.41 & 3.95 & 0.55 & 5.9 \\
SPT-CL J0444-4352 & 71.17 & -43.87 & 3.69 & 0.53 & 5.0 \\
SPT-CL J2020-6314 & 305.03 & -63.24 & 3.74 & 0.54 & 5.2 \\
SPT-CL J0030-5213 & 7.53 & -52.22 & 3.62 & 0.53 & 5.5 \\
SPT-CL J0512-5139 & 78.16 & -51.66 & 4.11 & 0.57 & 6.6 \\
SPT-CL J0152-5303 & 28.23 & -53.05 & 4.70 & 0.61 & 2.0 \\
SPT-CL J2206-5807 & 331.66 & -58.13 & 4.51 & 0.60 & 3.6 \\
SPT-CL J0157-4007 & 29.45 & -40.13 & 4.12 & 0.57 & 5.4 \\
SPT-CL J0649-4510 & 102.45 & -45.17 & 3.85 & 0.55 & 6.0 \\
SPT-CL J0543-4250 & 85.94 & -42.84 & 4.85 & 0.62 & 2.0 \\
SPT-CL J2110-5244 & 317.55 & -52.75 & 4.56 & 0.61 & 1.7 \\
SPT-CL J0243-5930 & 40.86 & -59.51 & 4.92 & 0.64 & 3.0 \\
SPT-CL J0217-5014 & 34.27 & -50.24 & 3.52 & 0.54 & 3.0 \\
SPT-CL J0256-5617 & 44.10 & -56.30 & 4.83 & 0.63 & 2.5 \\
SPT-CL J2312-4621 & 348.06 & -46.35 & 4.68 & 0.63 & 1.3 \\
SPT-CL J2354-5633 & 358.71 & -56.55 & 3.73 & 0.56 & 1.8 \\
SPT-CL J2220-4534 & 335.08 & -45.58 & 4.77 & 0.64 & 4.1 \\
SPT-CL J0519-4248 & 79.85 & -42.81 & 3.71 & 0.57 & 2.2 \\
SPT-CL J0422-5140 & 65.59 & -51.68 & 4.01 & 0.59 & 4.3 \\
SPT-CL J2140-5331 & 325.03 & -53.52 & 3.66 & 0.57 & 3.5 \\  \hline  
%\end{longtable*}
\multicolumn{6}{l}{} \\[-1pt] 
\caption{ \label{tab:TableI} The TS values for  SPT-SZ Galaxy Clusters  along with their masses and redshifts. The clusters are arranged in the decreasing order of their $M_{500}/z^{2}$  values.  The clusters with $TS > 9$ corresponding to at least $3\sigma$ significance  are shown in bold text.} 
\end{longtable*}
\end{center}

%\bibliographystyle{plain} % Choose a bibliography style
%\bibliography{references} % Replace 'references' with the name of your .bib file without the .bib extension

\end{document}